\documentclass[twocolumn,notitlepage,prl,superscriptaddress,longbibliography]{revtex4-2}

\usepackage{amssymb}
\usepackage{bm}

\usepackage{amsfonts}
\usepackage{textcomp}
\usepackage{times}
\usepackage{graphicx}
\usepackage{float}
\usepackage{latexsym,amsmath,amssymb,bm,euscript}
\usepackage{color}
\usepackage{subfigure}
\usepackage{epstopdf}
\usepackage[colorlinks=true,linkcolor=blue,citecolor=blue]{hyperref}
\usepackage{soul}
\usepackage[normalem]{ulem}
\usepackage{mathrsfs}
\usepackage{amsmath}
\usepackage{xspace}
\usepackage{natbib}
\usepackage{ulem}
\usepackage{etoolbox}
\usepackage{tikz}
\usepackage{physics}
\usepackage{chemformula}
\usepackage{natbib}
\usepackage{tikz}
\usepackage{txfonts}
\usepackage{ulem}
\usepackage{xspace}
\usepackage{xfrac}
\usepackage{mathtools}

\usepackage{titlesec}
\usetikzlibrary{decorations.markings}
\usetikzlibrary{decorations.pathmorphing}
\usetikzlibrary{arrows.meta}
\tikzset{middlearrow/.style={
        decoration={markings,
        mark= at position 0.6 with {\arrow{#1}} ,
       },
        postaction={decorate}
   }
}
\tikzset{snake it/.style={decorate, decoration=snake, segment length=5}}

\newcommand{\mf}{{\text{MF}}}
\newcommand{\hc}{\text{H.c.}}
\newcommand{\bdsb}[1]{\boldsymbol{#1}}

\newcommand{\tgt}{Fe$_3$GeTe$_2$}

\newcommand{\nnij}{\langle i,j\rangle}
\newcommand{\im}{\text{Im}}

\newcommand{\fgt}{\text{Fe$_3$GeTe$_2$}}

\newcommand{\rev}[1]{{\color{black}{#1}}}

\graphicspath{{./Fig/}{}}

\begin{document}
\title{Magnon Damping Minimum and Logarithmic Scaling in a Kondo-Heisenberg Model}

\author{Yuan Gao}
\thanks{These authors contributed equally to this work.}
\affiliation{School of Physics, Beihang University, Beijing 100191, China}
\affiliation{CAS Key Laboratory of Theoretical Physics, Institute of Theoretical 
Physics, Chinese Academy of Sciences, Beijing 100190, China}

\author{Junsen Wang}
\thanks{These authors contributed equally to this work.}
\affiliation{Center of Materials Science and Optoelectronics Engineering,
College of Materials Science and Opto-electronic Technology, University 
of Chinese Academy of Sciences, Beijing 100049, China.}
\affiliation{CAS Key Laboratory of Theoretical Physics, Institute of Theoretical 
Physics, Chinese Academy of Sciences, Beijing 100190, China}

\author{Qiaoyi Li}
\affiliation{CAS Key Laboratory of Theoretical Physics, Institute of Theoretical 
Physics, Chinese Academy of Sciences, Beijing 100190, China}
\affiliation{School of Physics, Beihang University, Beijing 100191, China}

\author{Qing-Bo Yan}
\email{yan@ucas.ac.cn}
\affiliation{Center of Materials Science and Optoelectronics Engineering,
College of Materials Science and Opto-electronic Technology, University 
of Chinese Academy of Sciences, Beijing 100049, China.}
\affiliation{CAS Center for Excellence in Topological Quantum Computation, 
University of Chinese Academy of Sciences, Beijng 100190, China}

\author{Tao Shi}
\email{tshi@itp.ac.cn}
\affiliation{CAS Key Laboratory of Theoretical Physics, Institute of Theoretical 
Physics, Chinese Academy of Sciences, Beijing 100190, China}
\affiliation{CAS Center for Excellence in Topological Quantum Computation, 
University of Chinese Academy of Sciences, Beijng 100190, China}

\author{Wei Li}
\email{w.li@itp.ac.cn}
\affiliation{CAS Key Laboratory of Theoretical Physics, Institute of Theoretical 
Physics, Chinese Academy of Sciences, Beijing 100190, China}
\affiliation{CAS Center for Excellence in Topological Quantum Computation, 
University of Chinese Academy of Sciences, Beijng 100190, China}
\affiliation{Peng Huanwu Collaborative Center for Research and Education, 
Beihang University, Beijing 100191, China}

\begin{abstract} 
Recently, an anomalous temperature evolution of spin wave excitations 
has been observed in a van der Waals metallic ferromagnet Fe$_3$GeTe$_2$ (FGT) 
[S. Bao, \textit{et al.}, Phys. Rev. X \textbf{12}, 011022 (2022)], whose theoretical 
understanding yet remains elusive. Here we study the spin dynamics of a 
ferromagnetic Kondo-Heisenberg lattice model at finite temperature, and propose 
a mechanism of magnon damping that explains the intriguing experimental results. 
In particular, we find the magnon damping rate $\gamma(T)$ firstly decreases as 
temperature lowers, \rev{due to the reduced magnon-magnon scatterings.} It then 
reaches a minimum at $T_{\rm d}^*$, and rises up again following a logarithmic 
scaling $\gamma(T) \sim \ln{(T_0/T)}$ (with $T_0$ a constant) for $T < T_{\rm d}^*$, 
\rev{which can be attributed to electron-magnon scatterings of spin-flip type.} 
Moreover, we obtain the phase diagram containing the ferromagnetic and Kondo 
insulator phases by varying the Kondo coupling, which may be relevant for experiments 
on pressured FGT. \rev{The presence of a magnon damping minimum and 
logarithmic scaling at low temperature indicates the emergence of the Kondo 
effect reflected in the collective excitations of local moments in a Kondo lattice system.} 
\end{abstract}

\date{\today}
\maketitle

{\textit{Introduction.---}}
The van der Waals ferromagnetic (FM) metal Fe$_3$GeTe$_2$ (FGT) 
has raised great research interest due to its tunable high-temperature
FM order~\cite{Verchenko2015,Deng2018} and anomalous Hall effect 
with a topological origin~\cite{Kim2018}, placing it a potential spintronic 
material~\cite{Fei2018,Deng2018,Wang2019SA}. 
Besides, FGT also raises fundamental research interest due to 
its intriguing thermodynamic, dynamical, and transport properties
\cite{Zhu2016DMFT,Zhang2018,Zhao2021KHM,Feng2022CPL,
Bao2022PRX,Goutam2022,Calder2022}. In particular, FGT has been 
proposed to exhibit intriguing Kondo-lattice behaviors
\cite{Zhu2016DMFT,Zhang2018,Zhao2021KHM,Feng2022CPL}, 
including the enhancement of Fermi surface volume, large effective 
electron mass~\cite{Zhang2018}, Fano resonance~\cite{Zhang2018},
and the presence of Kondo holes in the vicinity of Fe deficiencies
\cite{Zhao2021KHM}, etc. These experimental progresses point to 
the dichotomy of locality and itinerancy of the 3$d$ electrons in this 
FM metal~\cite{Zhang2018,Zhao2021KHM,Bao2022PRX,
Calder2022}.

Very recently, an anomalous damping of spin wave in FGT has been 
observed with inelastic neutron scattering (INS) measurements
\cite{Bao2022PRX}. The magnon dispersions are found to be more 
diffusive at lower temperature (e.g., 4~K) than those at higher values 
(e.g., 100~K). This is quite unusual for conventional ferromagnets
\cite{Slater1936,Stoner1947}, where the magnon excitations 
are expected to be more coherent as temperature decreases. The 
spontaneous magnon decay has been witnessed at low temperature 
for highly frustrated antiferromagnets~\cite{Chernyshev2009,Zhitomirsky2013,
Smit2020}, however, for such 2D ferromagnet the INS results with magnon 
damping remain to be understood. \rev{Moreover, latest INS experiments 
reveal a damping minimum and logarithmic scaling at low temperature
\cite{Bao2023}, which thus urgently calls for systematic theoretical 
studies on this 2D van de Waals ferromagnet.}

To address the anomalous magnon damping, here we consider a 
ferromagnetic Kondo-Heisenberg (FMKH) model for FGT [see 
Eq.~(\ref{Eq:Ham})] and study the temperature evolution of spin 
dynamics. The FMKH model consists of both itinerant electrons 
and local moments, \rev{due to the dual nature of $d$-orbital magnetism
in the compound.} With state-of-the-art tensor renormalization 
group (TRG) and perturbative field-theoretical calculations, 
we find the anomalous magnon damping in the FMKH model. 
In the FM phase of the model, itinerant $d$-electron scatters 
the collective magnon excitation of local moments via the Kondo 
coupling, which is renormalized in analogous to the original 
Kondo effect~\cite{Kondo1964,Anderson1970}. It gives rise to 
the anomalous decay of magnon characterized by a minimum in 
damping rate $\gamma(T)$ at $T_{\rm d}^*$ and a logarithmic 
temperature dependence $\gamma \sim \ln(T_0/T)$ for $T < T_{\rm d}^*$. 
Moreover, in the large Kondo coupling regime ($J_K \gg 1$), 
the conduction electrons can dynamically screen the local 
moments by forming Kondo singlets, and in such a Kondo 
insulator (KI) phase FM order vanishes while a coherent triplon 
wave forms. The FM and KI phases are \rev{separated by a
quantum critical point (QCP)}~\cite{Alexander2013,
Shen2020Nature,Kirkpatrick2020} that may be relevant 
for high-pressure experiments on FGT. 

\begin{figure}[!htbp]
\includegraphics[angle=0,width=1\linewidth]{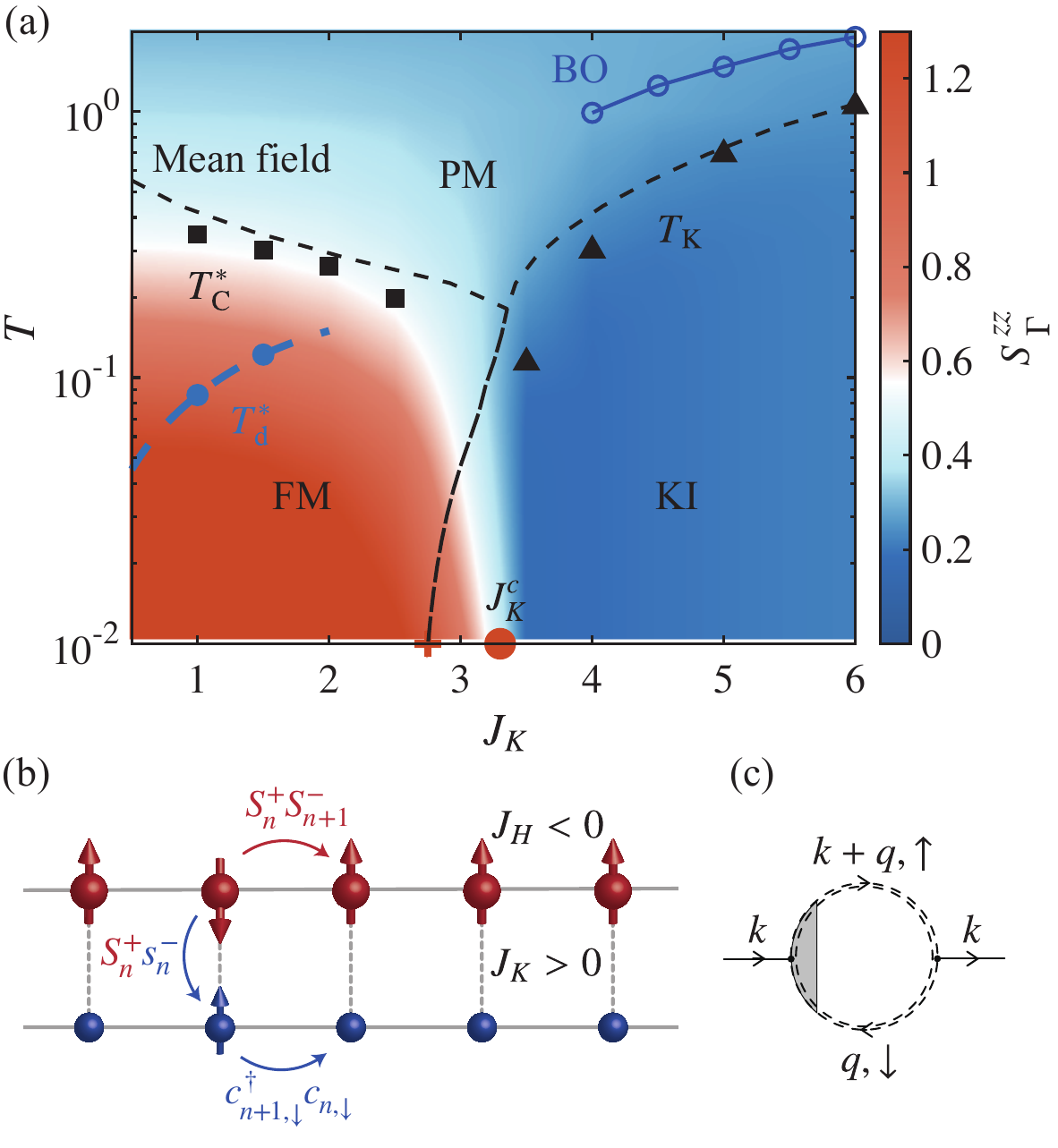}
\caption{(a) The finite-temperature phase diagram of FMKH model. 
The contour background is the spin structure factor $S^{zz}_{k=\Gamma}$. 
The black dashed line is the phase boundary obtained by mean-field 
calculations which separates the paramagnetic (PM), ferromagnetic 
(FM) and Kondo insulator (KI) phase. In the ground state, the FM and 
KI phases are separated by a QCP at $J_K^c/|J_H| \approx 3.3$ indicated 
by the red circle obtained by DMRG, and the mean-field result shows a 
transition point at $\tilde{J}_K^c/|J_H| \approx 2.75$. The solid squares 
($T_{\rm C}^*$) and triangles ($T_{\rm K}$) are obtained by TRG. 
The blue hollow circle of $T_{\rm K}$ is obtained from the bond 
operator (BO) theory. $T_{\rm d}^*$ in the FM phase represents 
the characteristic temperature with the damping rate minimum. 
(b) Magnon damping mechanism. The spin flip dissipates into the 
itinerant electron host through the Kondo coupling.
(c) Feynman diagram of magnon self-energy due to electron-magnon 
scattering with the renormalized vertex responsible for the anomalous 
decay of magnon, \rev{where the double-dashed lines represent 
electron's full propagator}.
}
\label{Fig1}
\end{figure}

\textit{Model and methods.---}
The FMKH model considered in the present work reads
\begin{equation}
H =  - t  \sum_{n, \sigma} \left(c_{n+1,\sigma}^\dagger c_{n,\sigma} 
+ \hc\right) + J_H \sum_{n} {\bdsb S}_n \cdot {\bdsb S}_{n+1} + 
J_K \sum_{n} {\bdsb S}_n \cdot {\bdsb s}_n,
\label{Eq:Ham}
\end{equation}
where $c^\dagger_{n,\sigma}$ ($c_{n,\sigma}$) is the creation 
(annihilation) operator of the conduction $d$-electron at site $n$ 
with spin $\sigma \in \{\uparrow, \downarrow\}$, and $\bdsb S_n$ 
($\bdsb s_n$) represents the spin operator of local moment 
(conduction electron) [c.f., Fig.~\ref{Fig1}(b)].

\rev{In the numerical simulations, we compute} the thermodynamic 
and dynamical properties of 1D FMKH chain with the TRG methods. 
The equilibrium thermal density matrix and related thermodynamic 
properties can be obtained by the matrix product operator (MPO) 
based TRG methods~\cite{Li2011,SETTN,Chen2017,Chen2018,
tanTRG2023}, which are highly efficient and accurate for both 
quantum spin ~\cite{HanLiSpin2019,Chen2019,Dong2017,
HLi2020PRR,Li2020TMGO,Li2021NC,YuCPL2021} and fermion 
many-body systems ~\cite{Chen2019,tanTRG2023,Qu2022arXiv}. 
Given the MPO representation of the thermal density matrix, 
we generalize the tangent-space approach for real-time evolution
\cite{TDVP2011,TDVP2016,MPSManifold2014} to the mixed states 
and study the temperature evolution of the spectral density (c.f.
Supplemental Sec.~I~\cite{Supplemental}). The Kondo lattice models 
have been widely used in the studies of heavy-fermion systems
\cite{Doniach1977,Lobos2013,Komijani2018,Shen2020Nature}, 
and here we consider the half-filled case in the present study. 
The FM coupling strength $J_H=-1$ is taken as the energy unit 
and the hopping energy is set as $t/|J_H|= 2$ throughout the 
calculations. The intriguing interplay between the local moments 
and conduction electrons~\cite{Kondo1964,Matthias2007RMP,
Yang2008Nature} gives rise to a rich phase diagram. Beyond the 
1D Kondo lattices, the 2D FMKH models, particularly defined on 
the triangular lattice, are analyzed with field-theoretical approach
\cite{Supplemental}, where we find consistent conclusions. \rev{It 
suggests that the main conclusion of magnon damping minimum
and logarithmic low-temperature scaling in damping rate is independent
of specific lattice geometries or spatial dimensions.}

\textit{Finite-temperature phase diagram of the FMKH model.---}
We obtain the finite-temperature phase diagram of 1D FMKH 
model [Eq.~(\ref{Eq:Ham})] with the exponential TRG approach
\cite{Chen2017,Chen2018,Li2019}. The simulations are restricted
within the 2-leg ladder structure of size $N=L\times2$ with the 
length $L$ up to 36, where one leg forms a chain of local moments 
while the other represents itinerant electrons [c.f. Fig.~\ref{Fig1}(b)]. 
The retained bond dimension is $D = 1600$, which guarantees 
accurate results till low temperature $T/|J_H| \simeq 0.01$, with 
the data convergence always checked~\cite{Supplemental}. 
At low temperature, the TRG calculations are found to converge 
towards the ground-state density matrix renormalization group 
(DMRG) results.

\begin{figure*}[t]
\includegraphics[angle=0,width=1\linewidth]{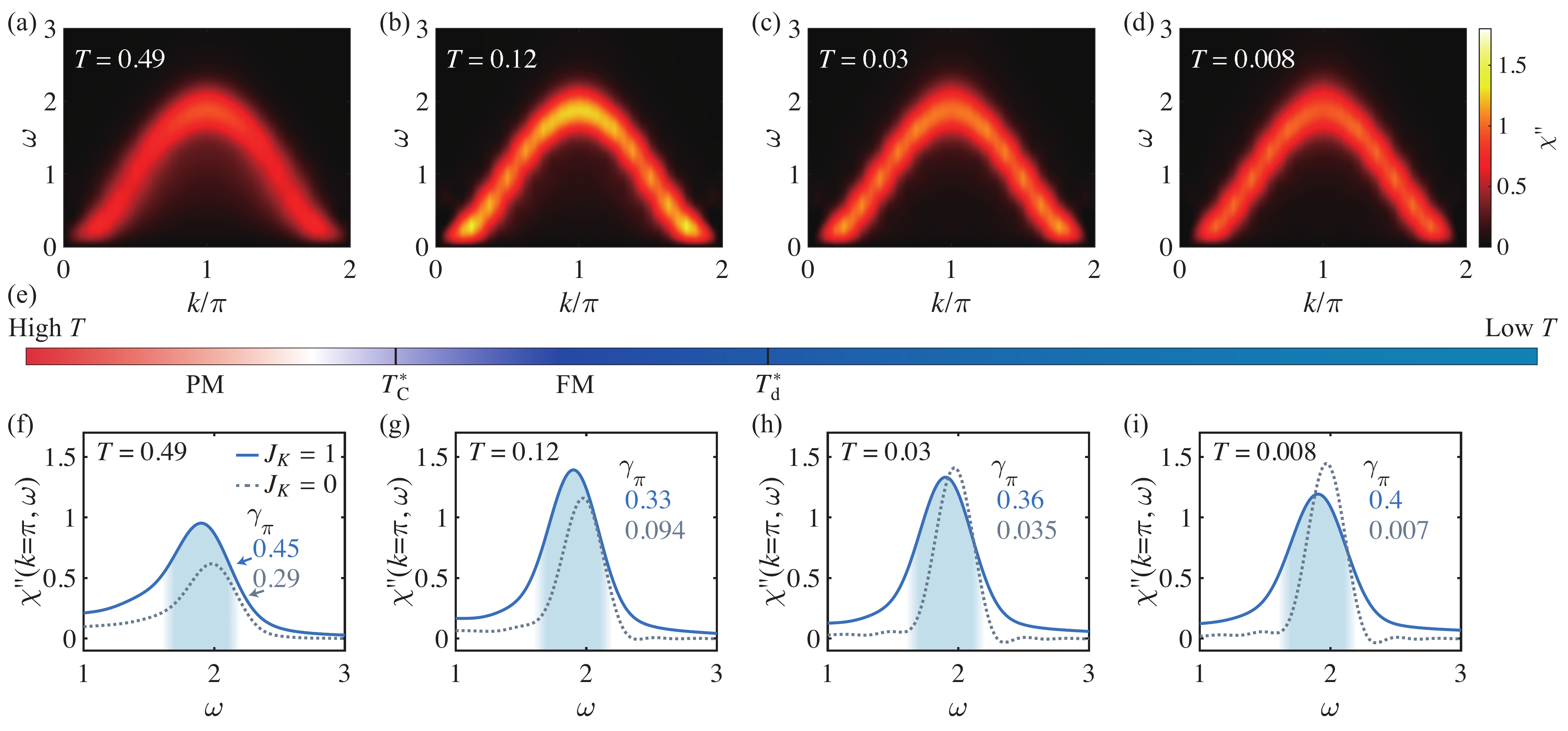}
\caption{ (a-d) show the contour plots of spectral density $\chi ''$ 
at various temperatures computed in the $N = 16 \times 2$ system
and with a retained bond dimension $D=1600$,
and (e) shows the corresponding phase diagram. 
(f-i) show the comparisons of $\chi''(k=\pi, \omega)$ between the 
Heisenberg case ($J_K=0$, calculated by ED) and Kondo-Heisenberg 
case ($J_K=1$, $D=4800$). The fitted damping rate $\gamma_\pi$ 
is shown beside the peak (blue for Kondo-Heisenberg case and 
gray for pure Heisenberg case).
}
\label{Fig2}
\end{figure*}

In Fig.~\ref{Fig1}(a) we show the obtained phase diagram of the FMKH 
model, with the contour plot of the spin structure factor $S^{zz}_k = 
\frac{1}{L} \sum_{m,n} \langle S_m^z S_n^z \rangle e^{{\rm i}(n-m)k}$ 
at $k=0$ (i.e., $\Gamma$ point) shown in the background. In the 
phase diagram we can recognize the existence of a ferromagnetic 
(FM, red) and a Kondo insulator (KI, blue) phase. At zero temperature,
the two phases are separated by a quantum critical point (QCP),
from which emanating the quantum critical regime at low temperature. 
The QCP located at $J_K^c \simeq 3.3$ is determined by the DMRG 
calculations, where we find an algebraic specific heat at low $T$
\cite{Supplemental}. As temperature elevates, the FM and KI phases 
evolve into the high-temperature paramagnetic (PM) regime. 

In companion with TRG numerics, we also perform mean-field calculations.
By introducing the FM order parameter $m_c = \expval{s_n^z}$ and 
$m_f = \expval{S_n^z}$ and the Kondo hybridization order parameter
\cite{zhang2000,li2010,liu2013} $V = \expval{c^\dagger_{n\uparrow} 
f_{n\uparrow} + f^\dagger_{n\downarrow} c_{n\downarrow}}$, where 
$f_{n, \sigma}$ is the pseudo-fermion operator for the local moment 
and $V$ characterizes the KI phase, we treat the FM and KI phases 
on an equal footing. The original Hamiltonian is then decoupled into a 
mean-field one that can be solved self-consistently~\cite{Supplemental}. 
The resulting phase boundary shown by the black dashed line in 
Fig.~\ref{Fig1}(a) is found to be in qualitative agreement with the 
TRG results. This is witnessed by the nice agreement between the 
mean-field dashed lines with the scatters ($T_{\rm C}^*$ and 
$T_{\rm K}$) determined from the peak location of the temperature 
derivatives of local moment correlations $-{d F_{nn}}/{d T}$ 
(with $F_{nn} = \frac{1}{L-1} \sum_n \langle {\bdsb S}_n 
\cdot {\bdsb S}_{n+1} \rangle$) and the local Kondo correlations 
$-{d K_{nn} }/{d T}$ (with $K_{nn} = -\frac{1}{L} \sum_n \langle 
{\bdsb S}_n \cdot {\bdsb s}_n \rangle$)~\cite{Supplemental}. 
In Fig.~\ref{Fig1}(a), we also represent the phase boundary determined 
by bond operator theory in KI phase as the blue circle
~\cite{Supplemental,sachdev1990,jurecka2001,eder2019}, 
where qualitative agreements are also observed.

\textit{Damping rate minimum and logarithmic scaling.---} 
Given the MPO representation of the thermal density matrix $\rho(\beta) 
= e^{-\beta H}$ ($\beta\equiv 1/T$), the dynamical properties at temperature 
$T$ can be simulated through a successive real-time evolution.
The spectral function $\chi''(k,\omega) \approx \int_0^{t_{\rm max}} 
dt~{\rm Im}[g(k,t)] \sin{(\omega t)} \cdot W({t}/{t_{\rm max}})$ is of 
central interest in studying the dynamical properties at finite temperature, 
where $W$ is the window function, $t_{\rm max}$ is the evolution time, 
and the (dynamical) correlation function is $g(k,t) = \langle A_k(t) 
A_k^\dagger \rangle_\beta$, with $A_k = \frac{1}{\sqrt{L}} \sum_n^L 
S_n^+ e^{{\rm i} k n}$ the Fourier transformed spin operator. 
Therefore, the problem resorts to the calculations of time-dependent 
correlation function $g(k,t)$, which we exploit the tangent-space 
approach to compute~\cite{TDVP2011,TDVP2016,MPSManifold2014}. 
In the real-time evolution calculations, we consider a $16\times2$ lattice 
and evolution time up to $t_{\rm max}/|J_H| = 20$, with $D = 4800$ bond 
states retained (see the data convergence in the Supplemental Sec.~I~C
~\cite{Supplemental}).

In Fig.~\ref{Fig2}(a-d), we show the spectral function $\chi''(\pi, \omega)$ of 
the FMKH model ($J_K=1$) where the line width firstly decreases and reaches 
its minimum at around $T^*_{\rm d}$, while it becomes broadened again as $T$
further lowers [c.f., the mini phase diagram Fig.~\ref{Fig2}(e)]. This is in sharp 
contrast to the pure Heisenberg ($J_K = 0$) case where the line width narrows 
monotonically as temperature decreases and shows a very clear dispersion 
at low $T$~\cite{Supplemental}.

\begin{figure}[t]
\includegraphics[angle=0,width=1\linewidth]{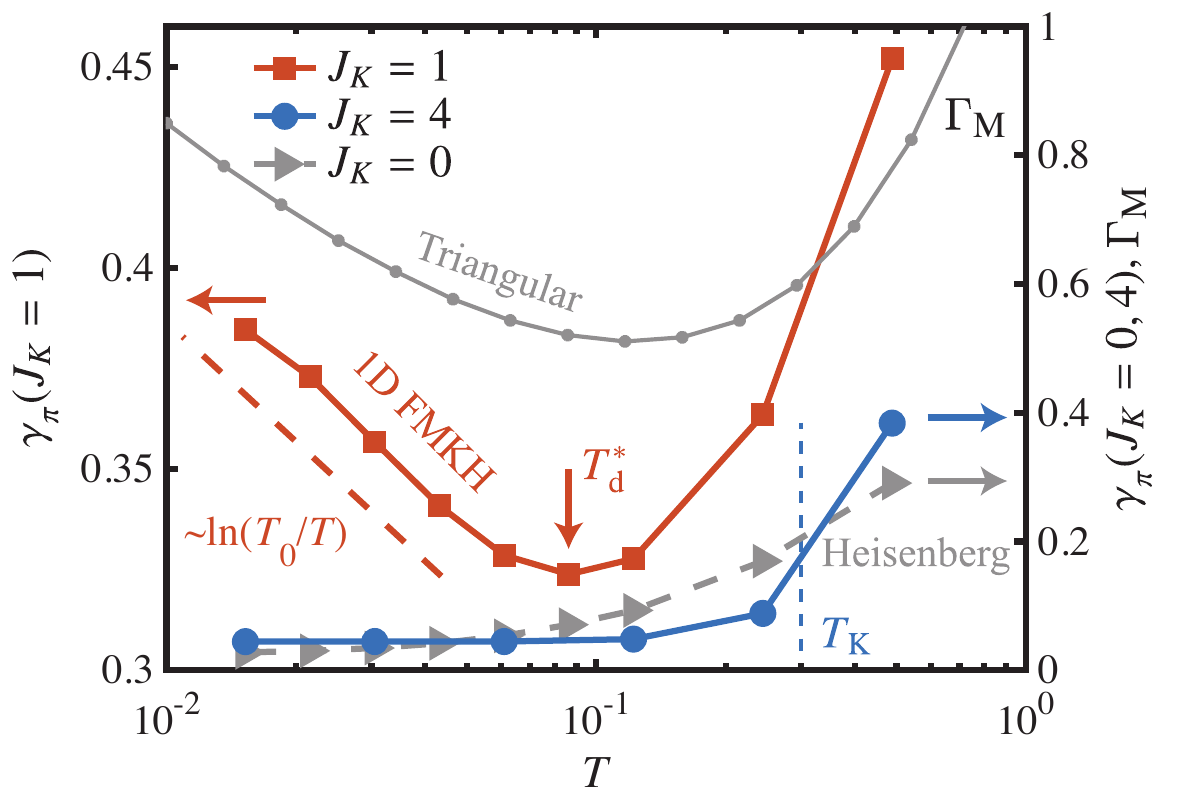}
\caption{Damping rate $\gamma_\pi$ for the FMKH model in the FM 
($J_K=1$) phase, as compared to that in the KI ($J_K=4$) phases 
and in pure Heisenberg model ($J_K=0$). In the FM phase of 1D 
FMKH model, the damping rate exhibits a minimum at the characteristic 
temperature $T_{\rm d}^*$, and below $T_{\rm d}^*$ it shows a logarithmic 
scaling $\gamma_{\pi} \sim \ln(T_0/T)$ with $T_0 \simeq 0.27$. 
The gray solid line ($\Gamma_{\rm M}$) is the damping rate of
triangular-lattice FMKH with $k = {\rm M} \equiv (0, \frac{\sqrt{3}}{3} \pi)$, 
$J_H = -1$, $t=2$, and $J_K = 10$ obtained via field-theoretical 
calculations (c.f. Supplemental Sec.~III~C~\cite{Supplemental}).
In contrast, the spin excitations become more coherent as $T$ 
lowers for either the Heisenberg or the KI case. The vertical blue 
dashed line shows the Kondo temperature $T_{\rm K} \simeq 0.3$ 
of the KI phase, as estimated in Fig.~\ref{Fig1}(a).}
\label{Fig3}
\end{figure}

It can be seen more clearly by plotting the $\chi''$ vs. $\omega$ 
at fixed $k=\pi$ [c.f., Fig.~\ref{Fig2}(f-i)], and compare the results 
to the pure Heisenberg ferromagnet. 
For the latter case, we find from the $\chi''(\omega)$ 
data that a more coherent peak develops as temperature lowers 
(i.e., absence of damping minimum); while for the $J_K=1$ case 
Kondo couplings introduce hybridization between local moments 
and the itinerant electron. With the spectral density results at $k = \pi$, 
we conduct a fitting with damped harmonic oscillator model 
\cite{andreas2002}, namely, $\chi''(k,\omega) \propto \frac{\gamma_k 
\omega E_k }{(\omega^2-E_k^2)^2 + (\gamma_k \omega)^2} \ast  
W(\omega)$ where $E_k$ is the magnon energy, $\gamma_k$ is 
the damping rate, and $W(\omega)$ is the convolution window 
function (c.f. Supplemental Sec.~II~B~\cite{Supplemental}).
Based on the fittings of spectral density at $k = \pi$ in Fig.~\ref{Fig2}(f-h), 
we collect the fitted damping {rate} $\gamma_\pi$ and show the results in 
Fig.~\ref{Fig3}, where $\gamma_\pi$ clearly exhibits a damping minimum 
at $T_{\rm d}^*$. Remarkably, the numerical simulations find a  
logarithmic scaling $\gamma_\pi(T) \sim \ln{T_0/T}$ below $T_{\rm d}^*$. 
Such a damping minimum and logarithmic scaling can be observed in 
various momentum $k$ and for different Kondo couplings $J_K < J_K^c$, 
i.e., it is a universal behavior due to the scattering between the collective 
excitations (magnons) of local moments and itinerant electrons 
(c.f. Supplemental Sec.~II~C~\cite{Supplemental}). 

As comparisons, we also show the damping rate $\gamma_\pi$ of the 
pure Heisenberg model ($J_K=0$) and that in the gapped KI phase with 
$J_K > J_K^c$~\cite{tsunetsugu1997}. As shown in Fig.~\ref{Fig3}, 
for either the pure Heisenberg or the KI case, $\gamma_\pi$ decreases 
monotonically as $T$ lowers and converges to a small value (i.e.,
long life time). In the KI phase, the local moments and the itinerant 
electrons form a singlet and the triplon wave moves coherently without 
scattering the electrons, as verified both by numerics (c.f., Supplemental 
Sec.~II~E~\cite{Supplemental}) and the bond-operator theory
\cite{sachdev1990,jurecka2001,eder2019} (c.f., Supplemental 
Sec.~V~\cite{Supplemental}).
Overall, the sharp contrasts in Fig.~\ref{Fig3} reveal the peculiarity in the 
magnon damping minimum and logarithmic scaling, which roots deeply
in the Kondo physics.

\begin{figure}[t]
\includegraphics[angle=0,width=1\linewidth]{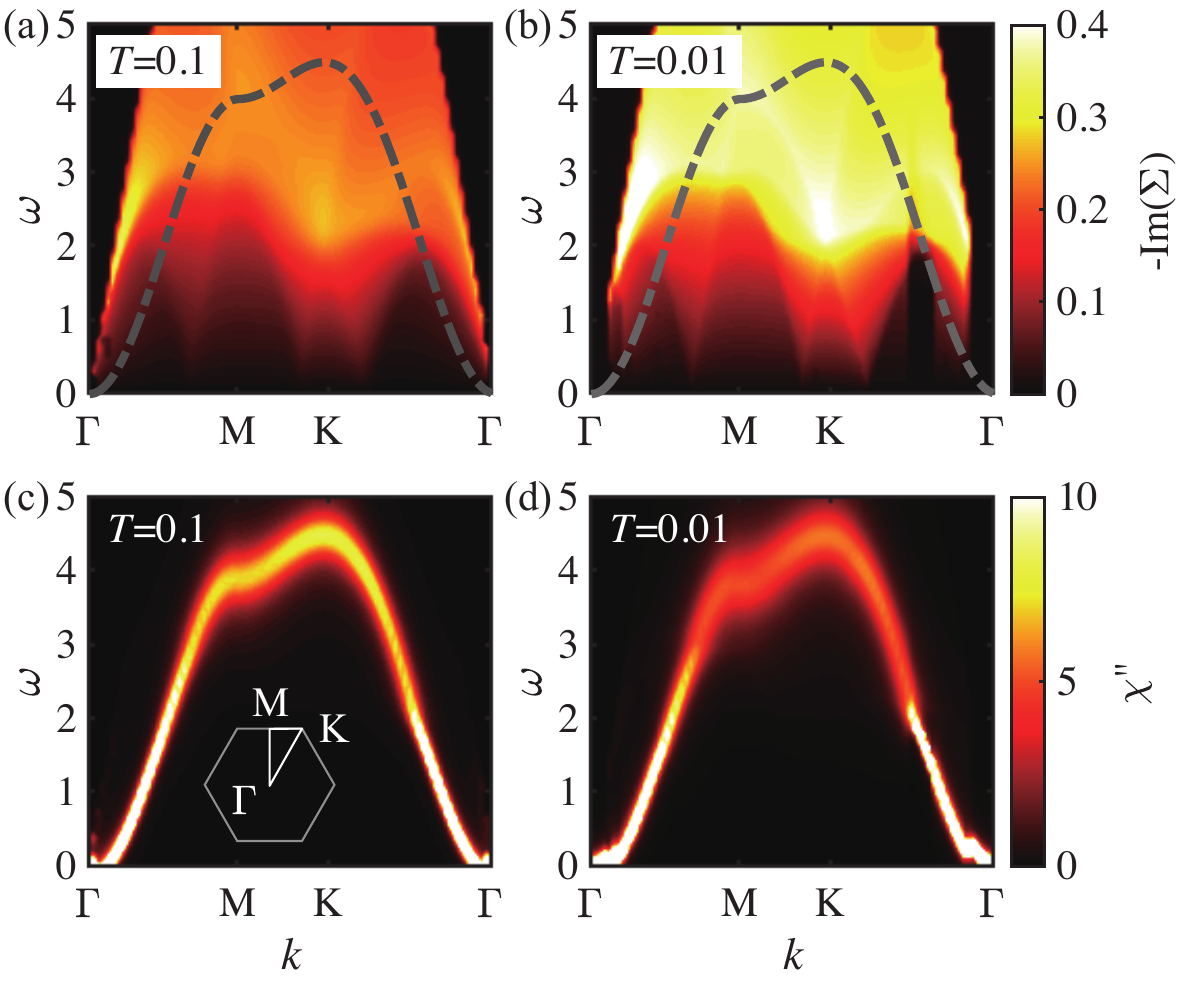}
\caption{(a, b) the contour plot of the magnon self-energy $-{\rm Im} (\Sigma)$ 
of the triangular-lattice FMKH model at various temperatures, where the gray 
dashed lines are the free magnon dispersion. The spectral density $\chi ''$ is 
shown in (c) and (d), with the $k$-path in the 1st Brillouin zone shown in the inset.
Other parameters used: $t=1$, $J_H=-1$ and $J_K=1$.}
\label{Fig4}
\end{figure}

\rev{
\textit{Field-theoretical analysis of the magnon decay.---} 
We now provide a physical picture to explain this anomalous magnon 
decay based on a field-theoretical analysis. In the weak-coupling limit, 
$J_K\ll |J_H|,t$, the spin part and fermion part can be dealt with separately first.
The former, via the Holstein-Primakoff mapping $S_i^z = S-n_i^a$ and 
$S_i^+ \approx \sqrt{2S} a_i$, can be approximated by free magnons; 
while the latter is simply the free fermions. In this same representation, 
the Kondo coupling term, treated as a small perturbation, approximately reads 
$J_K \sum_n {\bdsb S}_n \cdot {\bdsb s}_n \approx \frac{J_K}{2} \sum_i  
(a_i c^\dagger_{i\downarrow} c_{i\uparrow} + \mathrm{H.c.})$
\footnote{There is also a term involving density-density interaction between 
magnons and fermions that is less relevant in the current discussion.
As elaborated in more details in the Supplemental Materials,
it is just a vanishing Hartree contribution in the simplest approximation.}.
Thus the magnon self-energy, according to the Dyson equation 
\cite{Fetter2012}, is given diagrammatically in Fig.~\ref{Fig1}(c).
In sharp contrast to the usual electron-phonon interaction where the 
vertex correction is negligible due to Migdal theorem~\cite{Migdal1958}, 
here the vertex correction [shown as a gray area in Fig.~\ref{Fig1}(c)] 
due to the Kondo coupling $J_K$ is crucial~\cite{Kondo1964}.
By incorporating this renormalization effect into the calculation of the 
magnon self-energy, a $\log(1/T)$-like temperature dependence of 
self-energy at low-$T$ is manifested (c.f. Supplemental Sec.~III~A
\cite{Supplemental}). In Fig.~\ref{Fig3}, for triangular-lattice model 
we incorporate magnon-magnon 
interactions that leads to a power-law divergence also at high-$T$, 
and find the magnon damping rate, proportional to the imaginary part 
of the self-energy therefore exhibits a minimum at an intermediate 
temperature and a logarithmic scaling at low temperatures [c.f. more
details in Supplemental Sec.~III~C~\cite{Supplemental}].

Moreover, in Fig.~\ref{Fig4}(a-b), we show imaginary part of the 
renormalized magnon self-energy (omitting magnon-magnon interactions 
for simplicity) at relatively high ($T=0.1$) and low ($T=0.01$) temperatures,
where $-\Im(\Sigma)$ increases as temperature lowers. We also plot the 
free magnon dispersion as the gray dash-dotted lines, whose large $|\bdsb k|$ 
part is found to ``merge'' into the regime of large $-\Im(\Sigma)$ values. 
Consequently, in Fig.~\ref{Fig4}(c-d) we find the corresponding renormalized 
magnon spectral density $\chi''(\omega) = -\Im [G_{k}({\rm i} \omega_n 
\rightarrow \omega+{\rm i}0^+)]$ suffers heavier damping effect near the 
Brillouin zone boundary. This property, which is in agreement with recent
INS experiments~\cite{Bao2023} and also shared by other lattice geometries, 
has a kinematic origin --- There are many more damping channels for large 
$|\bdsb k|$. In Fig.~\ref{Fig4}(c-d), we plot the magnon spectral density at 
these two temperatures, and indeed find the dispersions get more blurred 
at a lower temperature $T=0.01$.
}

\textit{Discussion and Outlook.---} 
In this work, we study the finite-temperature dynamics of FMKH model 
with TRG calculations for 1D chain, with conclusions further supported
by field-theoretical analysis for 2D lattice. A magnon damping minimum 
and universal logarithmic scaling at low temperature are revealed, which
are ascribed to the scatterings between magnons and the itinerant 
electrons via the Kondo coupling. The logarithmical divergence
of the effective Kondo coupling in the renormalization process (i.e., 
vertex correction) is analogous to the renowned Kondo effect. Our 
results well explain, from a theoretical point of view, the INS 
measurements on FGT~\cite{Bao2022PRX,Bao2023}.

Based on the 1D model calculations, we uncover a QCP~\cite{Si2001,
Alexander2013,Komijani2018,Shen2020Nature} occurring at 
$J_K^c\simeq 3.3$ between the FM and KI phases in the phase diagram. 
We note in a recent experiment the Curie temperature of FGT compound 
decreases from $T_{\rm c}\simeq200$~K to 100~K as pressure increases, 
with local moments also prominently suppressed~\cite{Wang2019}. 
The high sensitivity of Curie temperature of the FM order in FGT 
upon compression is consistent with present FMKH model study. 
Further experimental investigations on the pressurized FGT are thus
called for according to our theoretical study here. 

\begin{acknowledgments}
\textit{Acknowledgments.---}
The authors are indebted to Song Bao, Zhao-Yang Dong, Xiao-Tian Zhang, Wei Zheng, Kun Chen, Jianxin Li,
and Jinsheng Wen for helpful discussions. This work was supported by 
the National Natural 
Science Foundation of China (Grant Nos. 12222412, 11974036, 11834014, 
and 12047503), the Fundamental Research Funds for the Central 
Universities, and CAS Project for Young Scientists in Basic Research 
(Grant No.~YSBR-057). We thank the HPC-ITP for the technical support 
and generous allocation of CPU time.
\end{acknowledgments}
\bibliography{KondoRef}
%
\newpage
\clearpage
\onecolumngrid
\mbox{}
\begin{center}
\textbf{\large Supplemental Materials: \\Magnon Damping Minimum and Logarithmic 
Scaling in a Kondo-Heisenberg Ferromagnet}\\

Gao \textit{et al}.
\end{center}

\date{\today}

\setcounter{section}{0}
\setcounter{figure}{0}
\setcounter{equation}{0}
\renewcommand{\theequation}{S\arabic{equation}}
\renewcommand{\thefigure}{S\arabic{figure}}
\setcounter{secnumdepth}{3}

\section{Calculations of the Dynamical Properties of
the Kondo-Heisenberg Model}
\subsection{Finite-temperature method: 
Exponential tensor renormalization group}
\begin{figure*}[htbp]
\includegraphics[angle=0,width=0.7\linewidth]{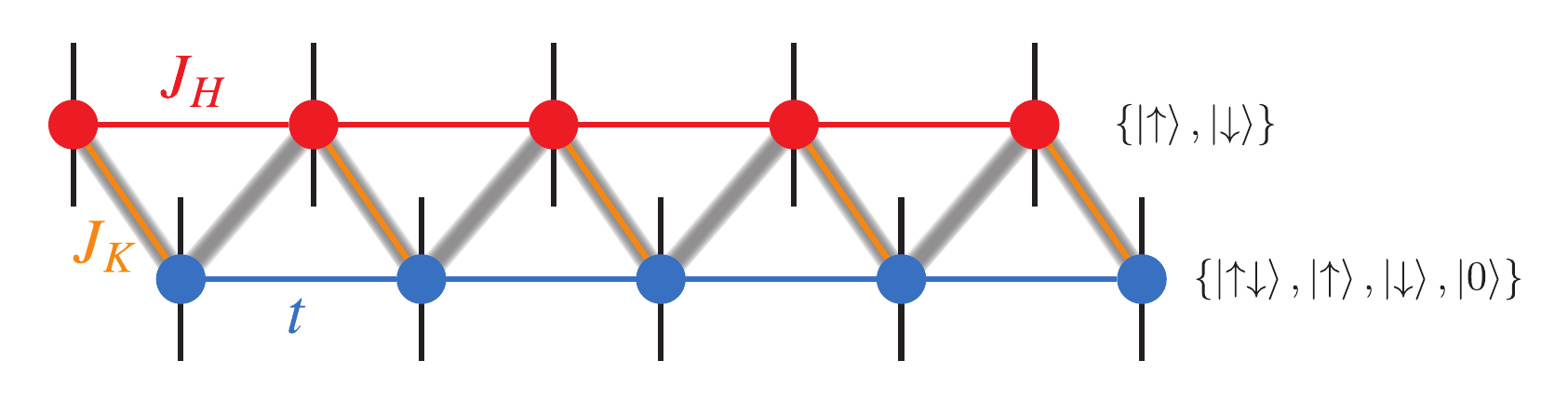}
\caption{The MPO representation of the Kondo-Heisenberg 
  model Hamiltonian. 
  The red and blue sites present the spin and fermion site respectively 
  with the corresponding local basis $\{\ket{\uparrow}, \ket{\downarrow}\}$ 
  and $\{\ket{\uparrow\downarrow}, \ket{\uparrow}, \ket{\downarrow}, 
  \ket{0}\}$. The red, blue and orange solid line represent the ferromagnetic
  exchange, hopping amplitude, and the Kondo coupling, respectively. The zigzag 
  gray line shows the mapping between the ladder lattice and the MPO 
  representation.}
\label{SFig:H}
\end{figure*}

The local Hilbert space of a fermion lattice site contains 4 states: the 
double occupied state $\ket{\uparrow\downarrow}$, single-occupied 
states $\ket{\uparrow}, \ket{\downarrow}$, and the empty site $\ket{0}$. 
Consider a Kondo-Heisenberg lattice model described by Eq. 1
in the main text, 
the double occupied state $\ket{\uparrow\downarrow}$ and empty site 
$\ket{0}$ of the spin (local moment) sites are projected out (c.f., 
Fig.~\ref{SFig:H}). To be specific, the Hilbert space is generated by 
the basis $(\{\ket{\uparrow}, \ket{\downarrow}\}_{s} \otimes 
\{\ket{\uparrow\downarrow}, \ket{\uparrow}, \ket{\downarrow}, 
\ket{0}\}_{f})^{\otimes N/2}$ where $s$ labels the spin (local moment) 
site while $f$ labels the fermion site, $N$ is the total number of spin
and fermion sites. The matrix product operator (MPO) representation of the Hamiltonian is 
shown in Fig.~\ref{SFig:H}.

We perform  exponential tensor renormalization group (XTRG)
simulations~\cite{Chen2018} of the finite-temperature properties, 
and to generate the finite-temperature density matrix for real-time 
simulations. In practical calculations, we start from the initial density 
matrix $\rho_0(\tau)$ of a very high temperature with 
$\tau = \mathcal{O}(10^{-4})$ ($T \equiv 1/ \tau$), represented in 
a MPO form via a series expansion 
\cite{Chen2017}
\begin{equation}
\rho_0(\tau) = e^{-\tau H} = \sum_{n=0}^{\infty} \frac{(-\tau H)^n}{n!} 
\simeq \sum_{n=0}^{N_{\rm cut}} \frac{(-\tau H)^n}{n!}.
\end{equation}
For a high temperature such as $\tau = 2.5\times 10^{-4}$, the series expansion 
saturates to machine precision with cutoff $N_{\rm cut}=6$-$7$.
Subsequently, by keeping squaring the density matrix repeatedly, i.e., 
$\rho_n(2^n\tau) \cdot \rho_n(2^n\tau) \to \rho_{n+1}(2^{n+1}\tau)$, 
we cool down the system along a logarithmic inverse-temperature 
grid $\tau \to 2\tau \to 2^2\tau \to ... \to 2^n\tau$. Such a cooling process 
reaches low temperature exponentially fast, and it reduces significantly  
the projection and truncation steps, rendering higher accuracy than
traditional linear evolution scheme, and thus it constitutes a very 
powerful thermal tensor network method for ultra-low temperature simulations.

\subsection{Real-time Evolution: TDVP for matrix product density operator}
In Heisenberg picture, the real-time evolution of a given operator 
$\mathcal{O}$ reads $\mathcal{O}(t) = e^{{\rm i} H t} \mathcal{O}_0 e^{-{ \rm i} H t}$ with 
the corresponding Heisenberg equation
\begin{equation}
{\rm i}  \frac{d}{dt} \mathcal{O} = -[H, \mathcal{O}] = \mathcal{O}H - H\mathcal{O}.
\label{EqS:HeiEq}
\end{equation}
In practice, we introduce an auxiliary parameter $t'$ such that $\mathcal{O}(t, t') = e^{{\rm i} H t} 
\mathcal{O}_0 e^{-{ \rm i} H t'}$, thus we can split Eq.~\ref{EqS:HeiEq} into two part, namely
\begin{subequations}
\label{EqS:Evo}	
\begin{align}
{\rm i}  \frac{d}{dt} \mathcal{O} &=- H\mathcal{O}, \\
{\rm i}  \frac{d}{dt'} \mathcal{O} &= \mathcal{O}H.
\end{align}
\end{subequations}

In the tensor-network language, we prepare the operator $\mathcal{O}$ and the 
Hamiltonian $H$ in the MPO form. And Eq.~\ref{EqS:Evo}
can be solved by the time-dependent variational principle (TDVP)~\cite{TDVP2011,
TDVP2016, MPSManifold2014} by introducing a Choi transformation~\cite{Choi1972}
\begin{subequations}
\begin{align}
{\rm i} \frac{d}{dt} |\mathcal{O}\rangle \rangle &= 
- H \otimes {I} |\mathcal{O}\rangle \rangle,\\
{\rm i} \frac{d}{dt'} |\mathcal{O}\rangle \rangle &= {I} \otimes H ^{\rm T} 
 |\mathcal{O}\rangle \rangle.
\end{align}
\end{subequations}
Noticing that $e^{{\rm i} t H\otimes I} = e^{{\rm i}t H} \otimes I$ and 
 $e^{-{\rm i} t' I\otimes H} = I\otimes e^{-{\rm i}t' H}$, we can solve Eq.~\ref{EqS:Evo}
 keeping the MPO form~\cite{tanTRG2023}.

In this article, we consider the correlation function $g(k, t)$ 
calculated at temperature $\beta \equiv 1/T$, i.e.,
\begin{equation}
g(k,t) = \langle A_k(t) A_k^\dagger \rangle_\beta 
=\frac{1}{Z} {\rm tr}[ e^{{\rm i} H t} e^{-\frac{\beta}{2} H} A_k e^{-{\rm i} H t} 
A_k^\dagger e^{-\frac{\beta}{2} H} ],
\end{equation}
where $Z$ is the partition function and 
\begin{equation}
A_k = \frac{1}{\sqrt{L}} \sum_{n=1}^L S_n^+ e^{{\rm i} k n},\quad k = \frac{2\pi m}{L},\quad m\in \{0,1,\cdots,L-1\}
\end{equation}
With the density matrix at inverse temperature ${\beta}/{2}$, 
the correlation function $g(k,t)$ can be obtained by performing 
real-time evolution of 
$e^{-\frac{\beta}{2} H} A_k$ following Fig.~\ref{SFig:tdvp}. 
In the present work, we perform calculations with retained bond 
dimension up to $4800$, $\delta t = 0.1$, and $t_{\rm max} = 20$.

\begin{figure*}[htbp]
\includegraphics[angle=0,width=0.7\linewidth]{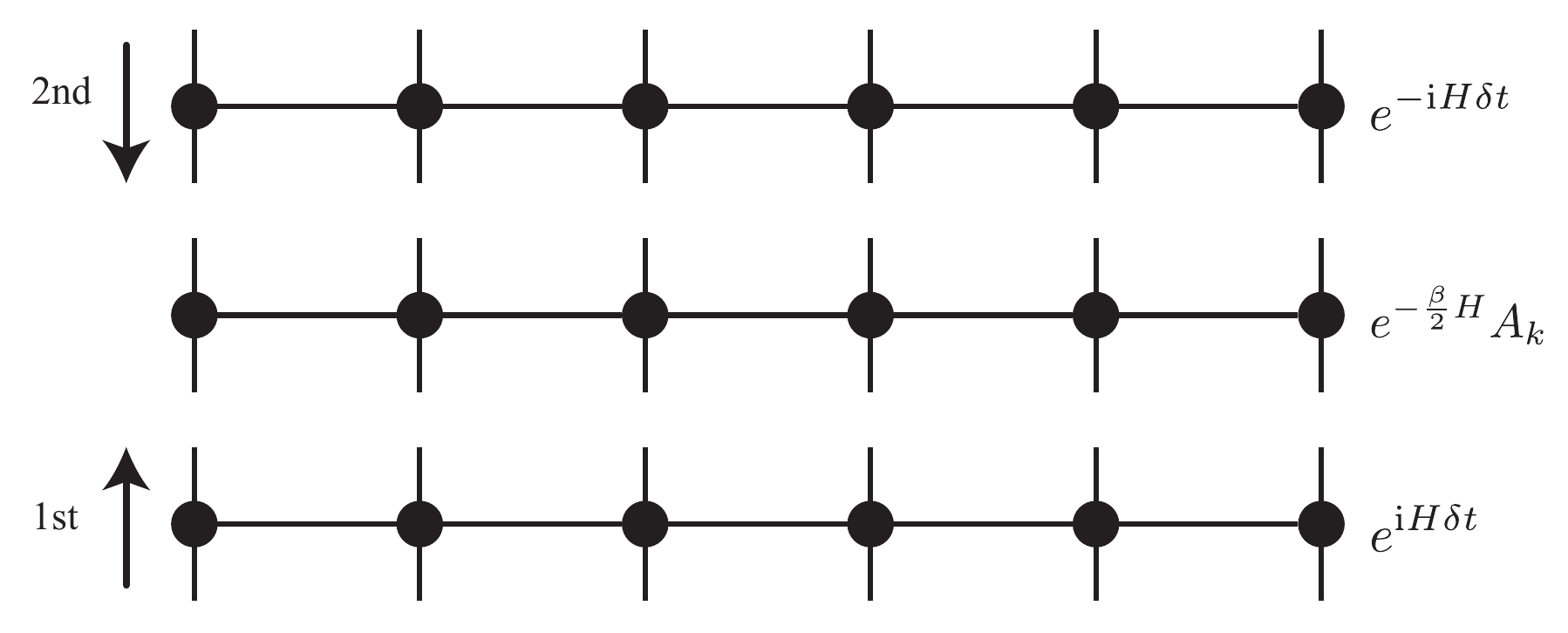}
\caption{The workflow of real-time evolution with a small step $\delta t$, 
  starting form the quenched finite temperature density matrix  
  $e^{-\frac{\beta}{2} H} A_k$. The evolution 
  operator $e^{{\rm i} H \delta t}$ and $e^{{-\rm i} H \delta t}$ are acted 
  on $e^{-\frac{\beta}{2} H} A_k$ following the order indicated
  by the arrows.}
\label{SFig:tdvp}
\end{figure*}

\subsection{Data convergence for real-time evolution}
\begin{figure*}[htbp]
\includegraphics[angle=0,width=0.95\linewidth]{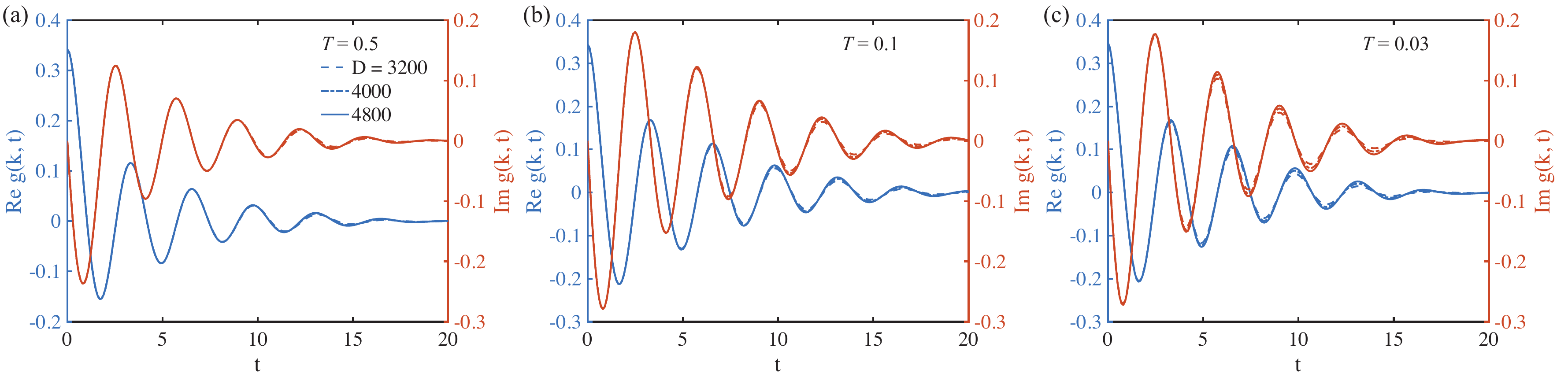}
\caption{Real (blue) and imaginary (red) part of the real-time 
correlation function with different bond dimensions for (a) $T=0.5$,
(b) $T=0.1$, and (c) $T=0.03$. The system size is $N=16\times 2$,
the hopping amplitude is $t = 2$, the Heisenberg exchange is 
$J_H = -1$ and the Kondo coupling $J_K=1$.}
\label{SFig:gkt}
\end{figure*}

In Fig.~\ref{SFig:gkt}, we show the real-time evolutions versus 
different bond dimension $D$, where we find the simulated results
have reached convergence for $D \gtrsim 4000$.

\section{Additional Data of the Spectral Density}

\subsection{Calculations of the spectral density}
The spectral function in a time translation invariant system
is defined as $\chi_{AB}''(t) = \frac{1}{2 \pi} \langle [A(t), B]\rangle$
and below we consider the spin excitation in the FMKH model, with $A=S^+$ and $B=A^\dagger = S^-$. 
The correlation function $g(k, t)$ is defined as
\begin{equation}
g(k,t) = \langle S^+_k(t) S_k^- \rangle_\beta 
= \frac{1}{Z}{\rm tr}[ e^{{\rm i} H t} e^{-\frac{\beta}{2} H} S^+_k e^{-{\rm i} H t} 
S_k^- e^{-\frac{\beta}{2} H} ] = g(k, -t)^\ast.
\end{equation}
Due to the spatial inversion symmetry, $g(k, t) = g(-k, t)$. Moreover, 
we have $\langle S^+(t) S^-\rangle_\beta = \langle S^-(t) S^+\rangle_\beta$ for the present system. 
Thus the spectral density reads
\begin{equation}
\chi''(k,t) =  \frac{1}{2 \pi} [g(k,t) -
g(k,t)^*] = \frac{{\rm i}}{\pi} {\rm Im}~ g(k,t),
\end{equation}
and
\begin{equation}
\chi''_{\rm def}(k, \omega) \equiv \int_{-\infty}^{+\infty} dt ~e^{{\rm i } t \omega} \frac{{\rm
i}}{\pi} {\rm Im}~ g(k,t)
=-\frac{2}{\pi} \int_{0}^{\infty} dt ~{\rm Im} ~g(k,t) \sin \omega t.
\end{equation}

Since the real-time simulation can only be performed for a finite 
period of time, the spectral function can be obtained approximately as
\begin{equation}
\chi''(k,\omega) \approx -a \int_0^{t_{\rm max}} dt ~{\rm Im} ~g(k,t) \sin \omega t \cdot 
W(\frac{t}{t_{\rm max}}),
\label{EqS:chipp}
\end{equation}
where $W(t)$ is a window function, and $a$ is a nonsignificant constant 
number (we set $a = 1$ in the follow part). In practice, the Hanning 
window defined below is chosen,
\begin{equation}
W(t) = \begin{cases}
\frac{1}{2} + \frac{1}{2} \cos(\pi t), &|t| \leq 1 \\
0, & |t| > 1.
\end{cases}
\end{equation}

\subsection{Kondo-Heisenberg model and the damped 
harmonic oscillator analysis}
We fit $\chi ''$ by the damped harmonic oscillator (DHO) model defined as 
\cite{andreas2002}
\begin{equation}
\chi''_{\rm def}(k,\omega) \propto \frac{\gamma_k 
\omega E_k }{(\omega^2-E_k^2)^2 + (\gamma_k \omega)^2}
\end{equation}
where $E_k$ is the magnon energy and $\gamma_k$ is the damping 
rate. Since the time window of real-time evolution is limited by 
tensor network calculations, we add a Hanning window for cut-off. In 
practice, the DHO fitting should take into account of the influences from 
the window.

\begin{figure*}[htbp]
  \includegraphics[angle=0,width=0.75\linewidth]{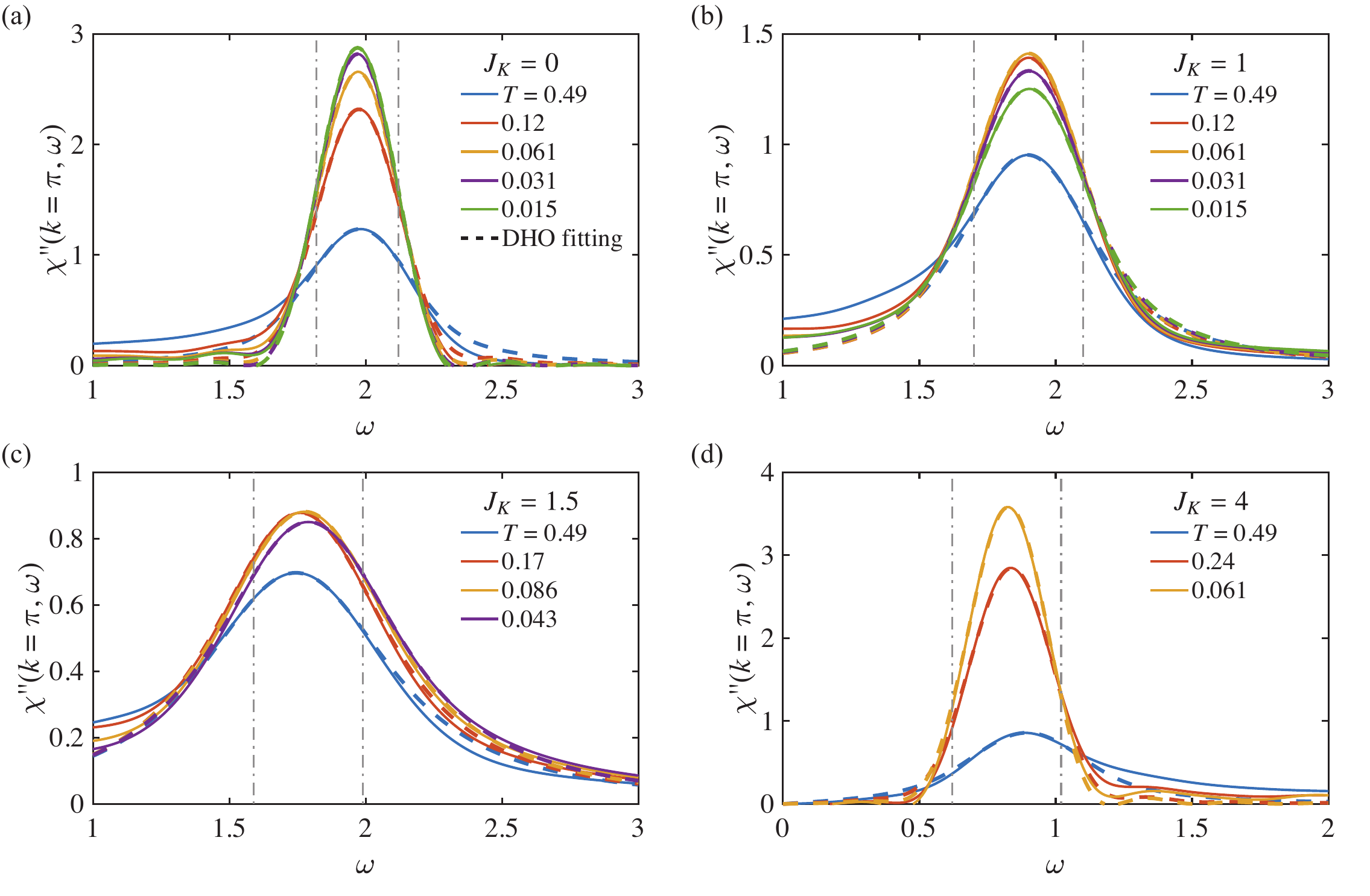}
  \caption{The DHO fittings of $\chi''(k,\omega)$ results of (a) Heisenberg
  model with Kondo coupling $J_K=0$ (ED), (b) $J_K=1$ ($D=4800$), 
  (c) $J_K=1.5$ ($D = 3200$), and (d) $J_K=4$ ($D=1600$). 
  The solid lines are the calculated $\chi''(k,\omega)$ results at different 
  temperatures, and the correspond DHO fittings are shown with dashed 
  lines and in the same color code. The fitting window is indicated by the 
  two gray lines. In (a) for the Heisenberg ferromagnet with $J_K=0$, we take 
  $J_H=-1$ and the calculations are performed on a $L=16$ chain. For the
  $J_K>0$ cases in (b,c,d) panels, instead we take $t = 2$, $J_H=-1$ and
  the system size is $N=16\times2$.
  }
  \label{SFig:Dampfit}
\end{figure*}

From Eq.~\ref{EqS:chipp}, one may immediately find that 
\begin{equation}
\chi'' (\omega) = \frac{1}{2\pi} \chi''_{\rm def}(\omega) \ast W(\omega).
\end{equation}
Thus the DHO fitting is
\begin{equation}
\chi'' (\omega) \propto \frac{\gamma_k \omega E_k }{(\omega^2-E_k^2)^2 + (\gamma_k \omega)^2} \ast W(\omega),
\end{equation}
where
$W(\omega) = \frac{\pi^2 \sin(t_{\rm max} \omega)}
{-\pi^2 \omega + t_{\rm max} \omega^3}$ is the fourier transform of  Hanning window.

The DHO fitting results are shown in Fig.~\ref{SFig:Dampfit} 
with the $\chi''(k,\omega)$ obtained by TDVP ($J_K>0$) and exact 
diagonalization ($J_K=0$). Within the fitting range, i.e. the gray line 
in Fig.~\ref{SFig:Dampfit}, the DHO fitting shows excellent 
agreement with $\chi''(k,\omega)$.

\subsection{Damping rate minima and logarithmic scaling for different  $J_K$}
\begin{figure*}[htbp]
  \includegraphics[angle=0,width=1\linewidth]{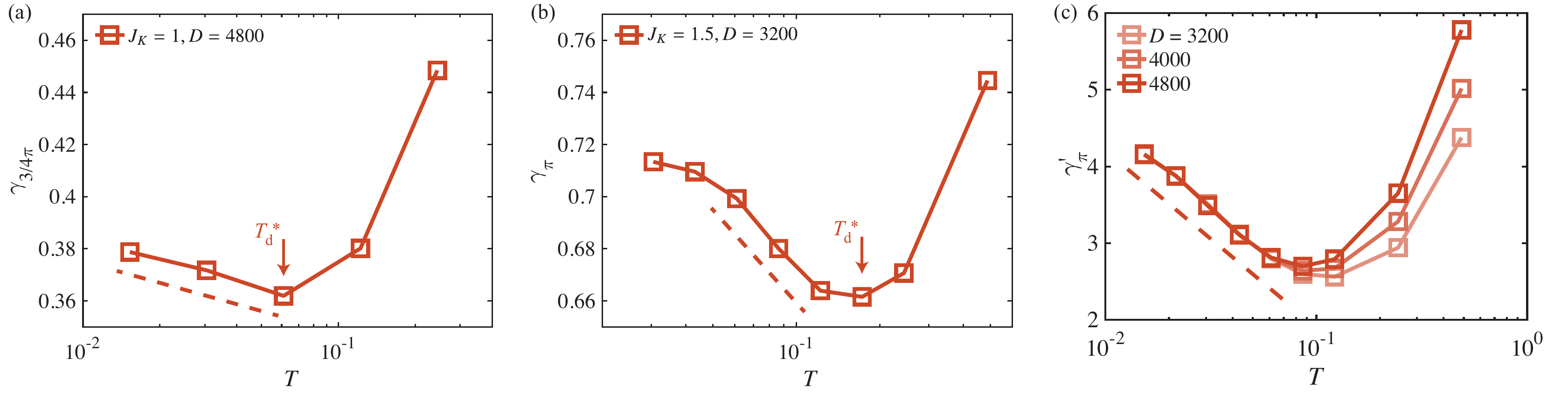}
  \caption{Supporting data of the damping coefficient $\gamma_k$ for 
  more coupling parameter $J_K$ and momentum $k$.
  (a) $J_K=1$ ($k=4\pi/3$) 
  and (b) $J_K=1.5$ ($k=\pi$), simulated on a $N=16\times2$ geometry. 
  $T_{\rm d}^*$ indicates the damping minimum of $\gamma_k$ and 
  the dashed line represents the logarithmic scaling.
  (c) shows the rescaled damping coefficient $\gamma'_\pi$ 
  with different bond dimension $D$, where the logarithmic scaling is found
  to be robust for various $D$.}
  \label{SFig:Damp}
\end{figure*}

The damping minimum and the logarithmic scaling can also be obtained 
with different momentum $k$ and Kondo coupling $J_K$ as shown in Fig.
\ref{SFig:Damp}. Since the damping coefficient is related to the magnon 
self-energy ${\rm Im} (\Sigma)$, the momentum $k<1$ is lesser influenced 
by the Kondo effect, i.e., the damping minimum is reached with a smaller 
temperature [see Fig.~\ref{SFig:Damp} (a)]. For $J_K =1.5$, we obtained the 
damping coefficient with $N=16\times 2$ and $D = 3200$. As shown in 
Fig.~\ref{SFig:Damp} (b), $\gamma_\pi$ shows similar behavior to the 
main text case. Increasing $J_K$, $T_{\rm d} ^ *$ also becomes higher.

To confirm the low-temperature scaling of the damping constant, we fit 
the damping coefficient $\gamma_\pi$ under different retaining bond dimension
by the model $a \log(b/T) + c$. The rescaled damping coefficient $\gamma'_\pi=
\frac{1}{a} (\gamma_\pi-c) - \log(b)$ is shown in Fig.~\ref{SFig:Damp}(c), and 
the low-temperature part shows exactly same behavior.

\subsection{Ferromagnetic Heisenberg chain: ED calculations}
\begin{figure*}[htbp]
\includegraphics[angle=0,width=1\linewidth]{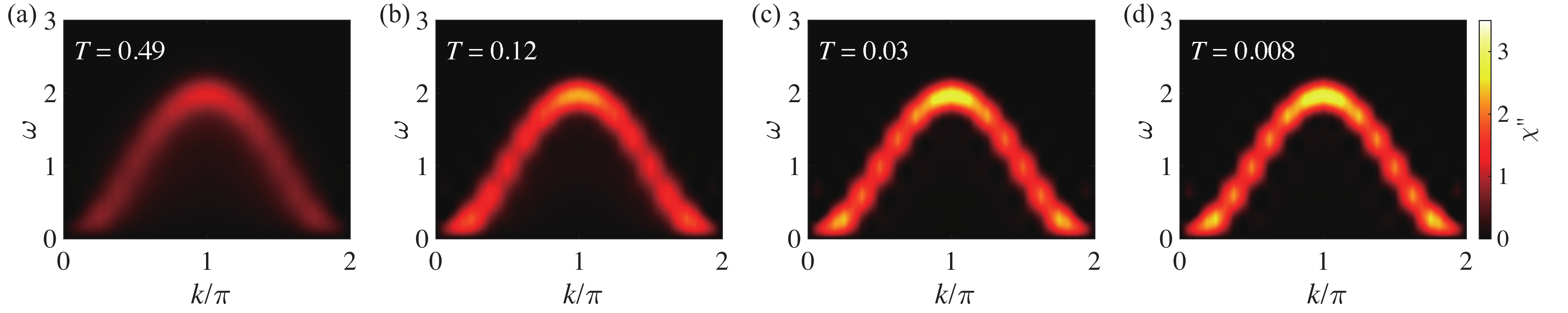}
\caption{Contour plot of spectral density $\chi''$ of a pure FM 
Heisenberg chain with $L=16$ at various temperatures.}
\label{SFig:HeiSkw}
\end{figure*}

As for the pure Heisenberg case ($J_K = 0$) with $L=16$, we obtain the dynamical
properties by exact diagonalization (ED) calculations. The contour plots 
of the spectral density are shown in Fig.~\ref{SFig:HeiSkw}, where the 
dispersion line shape becomes clearer as temperature lowers.

\subsection{Coherent triplon excitations in the Kondo insulator phase}
\begin{figure*}[htbp]
\includegraphics[angle=0,width=0.7\linewidth]{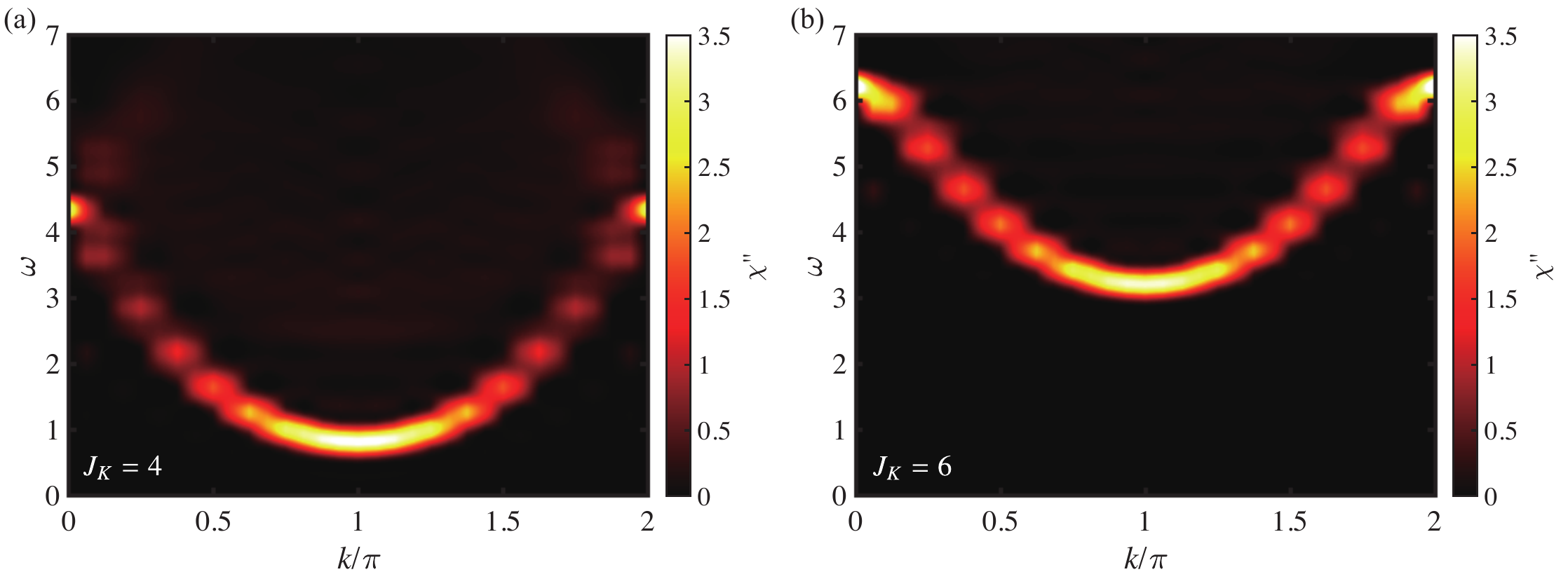}
\caption{Triplon dispersion for (a) $J_K=4$ and (b) $J_K=6$, with $D=1600$, 
$N=16\times2$ and $T=0.03$.}
\label{SFig:Triplon}
\end{figure*}

For sufficiently large Kondo coupling $J_K$, the system enters the Kondo 
insulator phase with coherent triplon waves. As shown in Fig.~\ref{SFig:Triplon}, 
we calculate the spin excitation dispersions with tensor network approach, 
and find the triplon excitation shows a square form dispersion with a minimum 
value at $k/\pi = 1$.

\section{Field Theoretical Analysis of the Magnon Damping Rate}

\subsection{Explicit evaluation of the bare polarization bubble diagram for the 1D chain}
\label{smsec:bubble}
The pair bubble diagram for the magnon self-energy, given in Fig.~1(c), \textit{using bare electron propagator and bare vertex}, reads
\begin{align}
  \Sigma_{\bdsb k} ({\rm i}\omega_n) &= \Sigma^{+-}_{\bdsb k} ({\rm i}\omega_n) \nonumber
\\
  & =\begin{tikzpicture}[baseline={([yshift=-.5ex]current bounding
box.center)},
  vertex/.style={anchor=base,circle,fill=black!25,minimum size=18pt,inner
sep=2pt}]
    \coordinate (l) at (0,0);
    \coordinate (r) at (1,0);
    \coordinate (ll) at (-0.2,0);
    \coordinate (rr) at (1.2,0);
    \draw[middlearrow={latex}] (l) to [out=90,in=90] (r);
    \draw[middlearrow={latex}] (r) to [out=-90,in=-90](l);
    \draw [snake it] (ll) -- (l);
    \draw [snake it] (rr) -- (r);
    \end{tikzpicture} \nonumber \\
   &= -\frac{J_K^2}{2} \int\frac{\mathrm d^dp}{(2\pi)^d} \frac{n_{c,\bdsb p+\bdsb
k}-n_{c,\bdsb p}}{{\rm i}\omega_n - 
   \epsilon_{c,\bdsb p+\bdsb k} + \epsilon_{c,\bdsb p}},\label{eq:ppbdiagram}
\end{align}
where $n_{c,\bdsb k} = 1/(e^{\beta \epsilon_{c,\bdsb k}}+1)$ is the Fermi
function.
For the 1D chain, after analytical continuation, ${\rm i}\omega_n \rightarrow \omega+{\rm i}0^+$, we have
\begin{align}
  \Sigma_{k}(\omega+{\rm i}0^+) &= -\frac{J_K^2}{2} 
  \int\frac{dp}{2\pi} \frac{n_{c,p+k} - n_{c,p}}{\omega - \epsilon_{c,p+k} + \epsilon_{c,p} + i0^+} \nonumber \\
  &= -\frac{J_K^2}{2} 
  \int\frac{dp}{2\pi} \frac{n_{c,p+k}- n_{c,p}}{\omega - \epsilon_{c,p+k} + \epsilon_{c,p}} 
  + {\rm i}
  \bigg[ \frac{\pi J_K^2}{2} \int\frac{dp}{2\pi}(n_{c,p+k} - n_{c,p}) \delta(\omega - \epsilon_{c,p+k} + \epsilon_{c,p}) \bigg].
\end{align}
From
\begin{align}
  \chi_k''(\omega) &= -\Im [G^R_{k}(\omega+{\rm i}0^+)] \nonumber \\ 
  &= -\Im \frac{1}{\omega - E_k - \Re \Sigma_k(\omega) - {\rm i}\Im\Sigma_k(\omega)}\nonumber\\
  &= \frac{-\Im \Sigma_k(\omega)}{[\omega-E_k -\Re\Sigma_k(\omega)]^2 + [\Im\Sigma_k(\omega)]^2},
\end{align}
the magnon damping rate is found to be
\begin{align}
  \gamma(\omega,k) &= \abs{\Im \Sigma_k(\omega)} \nonumber \\
  &= \abs{\frac{\pi K^2}{2} \int\frac{dp}{2\pi} (n_{c,p+k} - n_{c,p}) \delta(\omega - \epsilon_{c,p+k} + \epsilon_{c,p})} \nonumber\\
  &= \frac{J_K^2}{8t} \sum_{p_0} \abs{\frac{n_{c,p_0+k}-n_{c,p_0}}{\sin(p_0+k) - \sin p_0}} \nonumber\\
  &= \frac{J_K^2}{16t} \sum_{p_0} \abs{\frac{n_{c,p_0+k}-n_{c,p_0}}{\cos\frac{2p_0 + k}{2}\sin\frac{k}{2}}},\label{eq:gamma}
\end{align}
where the summation runs over all solutions of the following equation for $p_0\in [-\pi,\pi)$,
\begin{align}
  \omega &= \epsilon_{c,p_0+k} - \epsilon_{c,p_0}\nonumber\\
  &= 4t\sin\frac{2p_0+k}{2}\sin\frac{k}{2}.\label{eq:p0cond}
\end{align}
where $\epsilon_{c,k} = -2t \cos k$ is the dispersion of free fermion chain.
Focusing on the on-shell region, $\omega = \epsilon_k = \abs{J_H}(1-\cos k)$, Eq.~\eqref{eq:p0cond} always has two solutions given by $p_{0,1} = -\frac{k}{2} + \arccsc\bqty{\alpha\csc\frac{k}{2}}$ and $p_{0,2} = -\frac{k}{2} + \pi - \arccsc\bqty{\alpha\csc\frac{k}{2}}$ with $\alpha = 2t/|J_H|>1$ for the model parameters of \tgt. Limiting cases of the bare damping rate can be understood from the analytical expression Eq.~\eqref{eq:gamma}.
First, it is vanishingly small for small $k$.
Indeed, when expanded around $k=0$, it's straightforward to show that $\gamma\propto k$.
Secondly, it should reach the maximal value for $k$ near $\pi$, where the numerator becomes 
largest. Lastly, to the crudest approximation and as a reminiscence of the single-impurity Kondo 
problem, the Kondo coupling $J_K$ is renormalized as $J_K \rightarrow J_K^{\rm ren} = 
J_K[1 + J_K g(\epsilon_F) \log(\frac{\Lambda}{T})]$, where $g(\epsilon_F) = 1/(2\pi t)$ 
is the density of states per spin at the Fermi energy, and $\Lambda=4t$ is the band width.
When comparing with our numerical calculations given in Fig.~3, it is found that these model 
parameters should be further renormalized as $J_H=-0.07$ and $J_K=0.88$, which may 
be due to higher order corrections.

\begin{figure}[!t]
\includegraphics[width=0.9\linewidth]{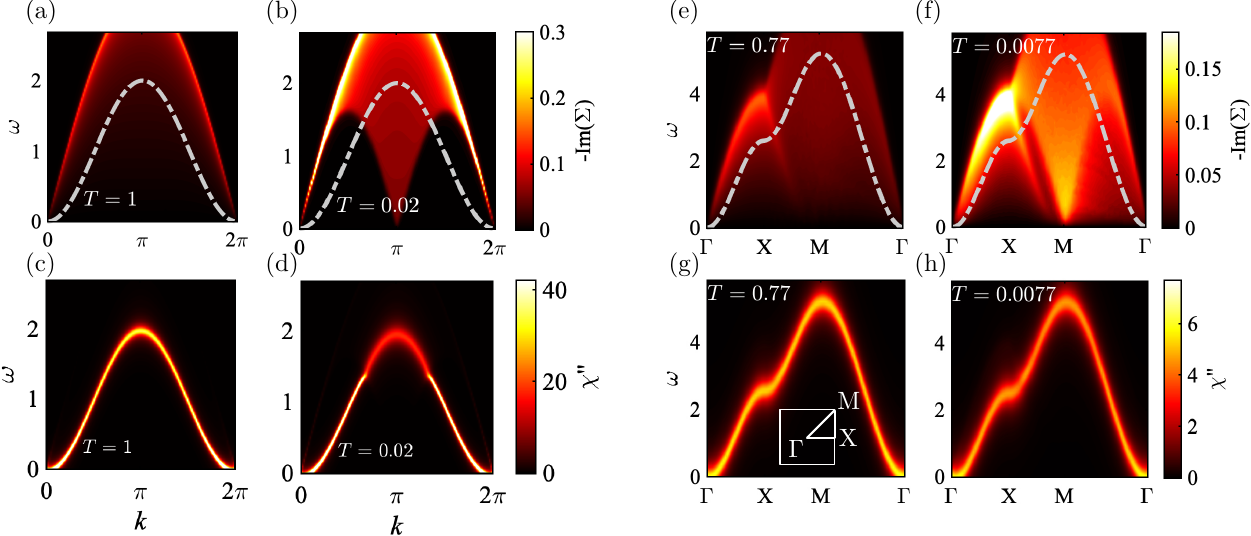}
\caption{First row: Contour plot of the imaginary part of magnon self-energy $\Im (\Sigma)$ for the Kondo-Heisenberg lattice model defined on the one-dimensional chain (a-b) and the square lattice (e-f), where the gray dash-dotted line is the corresponding free magnon dispersion. Second row: the corresponding magnon spectral density $\chi''=\Im G^R$, for the one-dimensional chain (c-d) and the square lattice (g-h) cases. The inset in (g) shows the momentum path used in the first Brillouin zone. Parameters used: $t/|J_H|=0.81$ and $J_K/|J_H| = 0.21$ for the one-dimensional chain case; $t/|J_H|=0.77$ and $J_K/|J_H|=0.077$ for the square lattice case.}
		\label{smfig:square_1d}
\end{figure}

We present numerical results of imaginary part of magnon self-energy in Fig.~\ref{smfig:square_1d}(a-b) and the corresponding spectral density in Fig.~\ref{smfig:square_1d}(c-d). It is found indeed that the imaginary part of the magnon self-energy increases at lower temperature and the magnon spectral density becomes more damped. Once again, we note that within this simple one-loop calculations, the magnon damping occurs only around $k\sim \pi$ and is absent for small $|k|$.

\subsection{The square lattice case}
For the FMKH model defined on the square lattice, with the primitive lattice vectors given by $\bdsb u_1 = (1, 0)$ and $\bdsb u_2 = (0, 1)$, the corresponding Brillouin zone is shown in the inset of Fig.~\ref{smfig:square_1d}(g). After using the Holstein-Primakoff transformation, the free magnon dispersion is $\epsilon_{\bdsb k} = J (2-\cos k_x - \cos k_y)$. Meanwhile, the free electron dispersion reads $\epsilon_{c,\bdsb k} = -2t(\cos k_x + \cos k_y)$. Hence the self-energy can be calculated numerically using Eq.~(3) in the main text, and results are shown in Fig.~\ref{smfig:square_1d}. Note at half-filling, the density of state of free fermion for the square lattice is divergent due to the presence of van Hove singularity, which is of course an artifact resulting from the approximation used in calculating the corresponding momentum integration. In practice, we simply regularize this density of state to a finite one in comparable to the one-dimensional case. The two main features, which are also shared for the one-dimensional case and the triangular lattice case, are that (1) the imaginary part of magnon self-energy increases at lower temperature, and the corresponding magnon spectral density is more damped; (2) damping phenomenon occurs primarily at large $|\bdsb k|$ and becomes absent for small $|\bdsb k|$. The second feature can easily be understood from the kinematics: the damping channel becomes negligible for small $|\bdsb k|$, therefore the magnon remains to be well-defined in that case.

\subsection{Magnon damping minimum for the triangular lattice case}
The calculations present in this section so far only produce $\log T$ scaling of magnon damping at low temperature. While in experiment it is found that there is also a magnon damping minimum at an intermediate temperature \cite{Bao2023}. In this subsection, we further consider magnon-magnon interactions beyond linear spin wave theory. It turns out that this additional contribution gives a power-law divergence at relatively high temperatures. Together with contributions from electron-magnon interactions, the magnon damping minimum found in experiments is therefore reproduced.

\begin{figure}[th!]
\centering
\includegraphics[width=\linewidth]{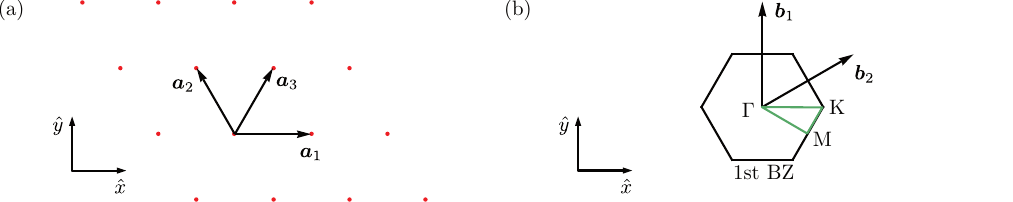}
\caption{(a) The triangular lattice is generated by linear combinations of the basis vectors $\bm{a}_1 = \pqty{1,0}$ and $\bm{a}_2=\pqty{-1/2,\sqrt{3}/2}$. (b) The corresponding reciprocal lattice is generated by the basis vectors $\bm{b}_1 = 2\pi\times(1,1/\sqrt{3})$ and $\bm{b}_2 = 2\pi\times(0,1/\sqrt{3})$. The first Brillouin zone (BZ) is shown as a hexagon, with the high symmetric path indicated by the green lines.}
\label{fig1}
\end{figure}

\begin{figure}[t]
\centering
\includegraphics[width=\linewidth]{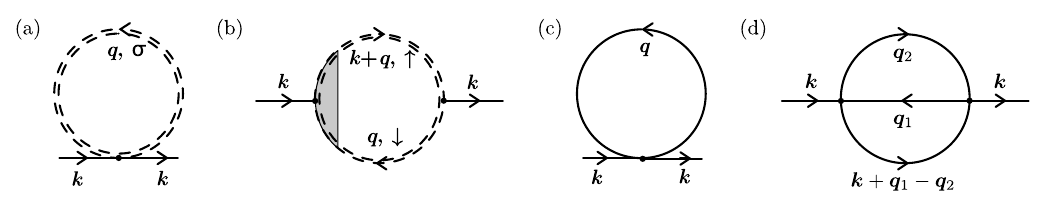}
\caption{Contributions to magnon self-energy due to electron-magnon interaction [(a) and (b)] and magnon-magnon interaction (up to zeroth order in $1/S$ and second order in $J$) [(c) and (d)]. Solid (dashed double) line represents free magnon (renormalized electron) propagators. \textit{For simplicity, we replace the renormalized electron propagator by the bare one for all our calculations.} The gray area in (b) indicates the usage of renormalized vertex. Here we also simply use the replacement $ J_K \rightarrow J_K^{(\text{ren})} =J_K [1+J_K g(\epsilon_F) \log(\frac{\Lambda}{T})]$, as in the case of single impurity Kondo model, which is responsible for the anomalous temperature dependence of magnon damping rate at low temperatures.}
\label{fig2}
\end{figure}

As the minimal model Hamiltonian for \tgt, the ferromagnetic Kondo-Heisenberg model on the triangular lattice reads $H = H_t + H_J + H_K$, where
\begin{align}
    H_t &= -t\sum_{\nnij} c^\dagger_{i,\sigma} c_{j,\sigma} + \hc \label{ht},\\
    H_J &= -J \sum_{\nnij} \bm{S}_i \cdot \bm{S}_j,\label{HJ}\\
    H_K &= J_K \sum_i \bm{S}_i \cdot \bm{s}_i. \label{HK}
\end{align}
Using the Holstein-Primakoff (HP) transformation,
\begin{subequations} \label{hpt}
\begin{align}
	S^+_i &= \sqrt{2S-a^\dagger_i a_i} a_i, \\
	S^-_i &= a^\dagger_i \sqrt{2S-a^\dagger_i a_i}, \\
	S^z_i &= S - a^\dagger_i a_i,
\end{align}
\end{subequations}
Equation~\eqref{HJ} can be rewritten as $H_J = H^{(-2)} + H^{(-1)} + H^{(0)} + \mathcal O\pqty{\frac{1}{S}}$. Here the superscript labels the corresponding order in $1/S$. Explicitly, we have
\begin{align}
	H^{(-2)} &= \pqty{\frac{1}{S}}^{-2} \times (-zNJ), \label{h2} \\
	H^{(-1)} &= \pqty{\frac{1}{S}}^{-1} \times 2zJ\sum_{\bm k} (1-\gamma_{\bm k}) a^\dagger_{\bm k} a_{\bm k}, \label{h1} \\
	H^{(0)} &= \pqty{\frac{1}{S}}^{0} \times\frac{J}{4} \sum_{\nnij} \pqty{a^\dagger_i a^\dagger_i a_i a_j + a^\dagger_j a^\dagger_j a_i a_j + a^\dagger_i a^\dagger_j a_i a_i + a^\dagger_i a^\dagger_j a_j a_j - 4 a^\dagger_i a^\dagger_j a_i a_j} \label{h0} \\
	&= \sum_{\{\bm k_i\}}{}^{'} V_4(\bm k_1, \bm k_2; \bm k_3, \bm k_4) a^\dagger_{\bm k_1} a^\dagger_{\bm k_2} a_{\bm k_3} a_{\bm k_4}. \label{h0p}
\end{align}
Equation~\eqref{h2} gives the classical ground state energy for a pure ferromagnetic Heisenberg model on the triangular lattice, with $z=3$ half the coordination number and $N$ the total lattice sites. Equation~\eqref{h1} gives the free magnon dispersion $\varepsilon_{\bm k} = 2zSJ(1-\gamma_{\bm k})$. For the triangular lattice, we have $\gamma_{\bm k} = \frac{1}{z}\sum_{l=1}^z \cos(\bm{a}_l \cdot \bm{k})$, with $\bm{a}_{1,2,3}$ shown in Fig.~\ref{fig1}(a). Equation~\eqref{h0} gives the interaction between magnons, where the interaction vertex explicitly reads $V_4(\bm k_1, \bm k_2; \bm k_3, \bm k_4) = \frac{J}{4	N} \pqty{\gamma_{\bm k_1} + \gamma_{\bm k_2} + \gamma_{\bm k_3} + \gamma_{\bm k_4} - 4\gamma_{\bm k_2 - \bm k_4}}$. The primed summation in Eq.~\eqref{h0p} means that the total momentum before and after scattering is conserved, i.e., $\bm k_1 + \bm k_2 = \bm k_3 + \bm k_4$. Using the same HP transformation, Equation~\eqref{HK} can be written as
\begin{equation}
    H = J_K \sum_i\bqty{ (S-n_i) (n^f_{i,\uparrow} - n^f_{i,\downarrow}) + \sqrt{\frac{S}{2}} (a_i c^\dagger_{i,\downarrow} c_{i,\uparrow} + a^\dagger_i c^\dagger_{i,\uparrow} c_{i,\downarrow})} + \mathcal O\pqty{\pqty{{\frac{1}{S}}}^{1/2}}.\label{hintjk}
\end{equation}
Based on Eq.~\eqref{h0p} and \eqref{hintjk}, all contributions to magnon self-energy up to the zeroth order in $1/S$ and second order in $J$ and $J_K$ are listed in Fig.~\ref{fig2}.

Figure~\ref{fig2}(a) is a Hartree contribution, which vanishes since $\langle n^f_{i,\uparrow} \rangle_0 = \langle n^f_{i,\downarrow} \rangle_0 $. Namely,
\begin{equation}
    \Sigma^{(a)}_R(\bm k) = 0.
\end{equation}

Figure~\ref{fig2}(b) explicitly reads $\Sigma^{(b)}(\bm k, i\omega) = J_K^2 \frac{S}{2} \chi_0(\bm k, i\omega)$, where the pair-bubble diagram is
\begin{align}
    \chi_0(\bm k, i\omega) &= \frac{1}{N}\sum_{\bm q} \frac{1}{\beta}\sum_{i\omega_1} \frac{1}{i\omega_1 - \xi_{\bm q}} \frac{1}{i\omega + i\omega_1 -\xi_{\bm k + \bm q}} \\
    &= \frac{1}{N}\sum_{\bm q} \frac{f_F(\xi_{\bm q}) - f_F(\xi_{\bm k+\bm q})}{i\omega - (\xi_{\bm k + \bm q} - \xi_{\bm q})},
\end{align}
and $\xi_{\bm k}=-2tz\gamma_{\bm k}$ is free electron's dispersion. After analytical continuation, and \textit{using renormalization of Kondo coupling} $J_K \rightarrow J_K^{\text{(ren)}} = J_K [1+J_K g(\epsilon_F) \log(\frac{\Lambda}{T})]$ reminiscent of the singlet-impurity Kondo problem~\cite{Kondo1964}, the retarded self-energy reads
\begin{equation}
\label{sigmab}
    \Sigma^{(b)}_R (\bm k, \omega) = J_K^2 \bqty{1 + J_K g(\epsilon_F) \log(\frac{\Lambda}{T})} \frac{S}{2N} \sum_{\bm q} \frac{f_F(\xi_{\bm q}) - f_F(\xi_{\bm k+\bm q})}{\omega - (\xi_{\bm k + \bm q} - \xi_{\bm q}) + i0^+}.
\end{equation}
Here $g(\epsilon_F)$ is the density of states per spin at the Fermi energy, and $\Lambda$ is the band width.

Figure~\ref{fig2}(c) is also a Hartree contribution. We replace the product of two operators by their expectation values in Eq.~\eqref{h0p}, which leads to four non-vanishing terms. After simplification, the upshot is simply a correction to the magnon dispersion relation,
\begin{equation}
    \Sigma^{(c)}_R(\bm k) = \sum_{\bm q} 2J(\gamma_{\bm q} + \gamma_{\bm k} - \gamma_0 - \gamma_{\bm k - \bm q}) f_B(\varepsilon_{\bm q}).
\end{equation}

Figure~\ref{fig2}(d) explicitly reads
\begin{align}
	\Sigma^{(d)}(\bm k, i\omega) &= \frac{J^2}{N^2\beta^2} \sum_{\{\bm q_i, i\omega_i\}} \frac{\abs{V_4 (\bm k, \bm q_1 ; \bm q_2, \bm k+\bm q_1 - \bm q_2)}^2}{(i\omega_1 - \varepsilon_{\bm q_2}) (i\omega_2 - \varepsilon_{\bm k + \bm q_1 - \bm q_2})[i(\omega_1 + \omega_2 - \omega) - \varepsilon_{\bm q_1}]}\\
	&= \frac{J^2}{N^2} \sum_{\{\bm q_i\}} \frac{\abs{V_4 (\bm k, \bm q_1 ; \bm q_2, \bm k+\bm q_1 - \bm q_2)}^2}{i\omega + \varepsilon_{\bm q_1} - \varepsilon_{\bm q_2} - \varepsilon_{\bm k + \bm q_1 - \bm q_2}} F_{\bm k, \bm q_1, \bm q_2}.
\end{align}
where $F_{\bm k, \bm q_1, \bm q_2} = \bqty{1+f_B(\varepsilon_{\bm q_2})} \bqty{1+f_B(\varepsilon_{\bm k + \bm q_1 - \bm q_2})}f_B(\varepsilon_{\bm q_1}) - f_B(\varepsilon_{\bm q_2}) f_B(\varepsilon_{\bm k + \bm q_1 - \bm q_2}) \bqty{1+f_B(\varepsilon_{\bm q_1})}$. After analytical continuation, the retarded self-energy reads
\begin{equation}
\label{sigmad}
    \Sigma^{(d)}_R(\bm k, \omega) = \frac{J^2}{N^2} \sum_{\{\bm q_i\}} \frac{\abs{V_4 (\bm k, \bm q_1 ; \bm q_2, \bm k+\bm q_1 - \bm q_2)}^2}{\omega + \varepsilon_{\bm q_1} - \varepsilon_{\bm q_2} - \varepsilon_{\bm k + \bm q_1 - \bm q_2} + i0^+} F_{\bm k, \bm q_1, \bm q_2}.
\end{equation}

Thus the total retarded self-energy reads $\Sigma_R = \Sigma_R^{(a)} + \Sigma_R^{(b)} + \Sigma_R^{(c)} + \Sigma_R^{(d)}$. And its imaginary part, relating to the magnon damping rate, reads
\begin{align}
    \Gamma(\bm k, \omega) &= \abs{\im \Sigma_R (\bm k,\omega)} \\
    &= \abs{\im \Sigma_R^{(b)}(\bm k, \omega) + \im \Sigma_R^{(d)}(\bm k, \omega)} \\
    &= | J_K^2 \bqty{1 + J_K g(\epsilon_F) \log(\frac{\Lambda}{T})} \frac{S\pi}{2N} \sum_{\bm q} \bqty{f_F(\xi_{\bm q}) - f_F(\xi_{\bm k+\bm q})} \delta(\omega - (\xi_{\bm k + \bm q} - \xi_{\bm q})) \nonumber \\
    &\quad + \frac{\pi J^2}{N^2} \sum_{\{\bm q_i\}} \abs{V_4 (\bm k, \bm q_1 ; \bm q_2, \bm k+\bm q_1 - \bm q_2)}^2 F_{\bm k, \bm q_1, \bm q_2} \delta(\omega + \varepsilon_{\bm q_1} - \varepsilon_{\bm q_2} - \varepsilon_{\bm k + \bm q_1 - \bm q_2})|.
\end{align}
There are two contributions: one results from the Kondo coupling, $\Sigma^{(b)}_R$, and the other is due to magnon-magnon interactions, $\Sigma^{(d)}_R$. Each of them is an integration, and the corresponding integrand consists of two parts, the scattering coefficient and the delta function. The former gives temperature dependence and overall relative strength among different momenta; while the latter provides the kinematic constraint. To understand this kinematic constraint more carefully, we define the scattering density of states~\cite{pershoguba2018} for each contribution,
\begin{align}
    K^{(b)}(\bm k) &= \frac{1}{N}\sum_{\bm q} \delta(\omega - (\xi_{\bm k+\bm q} - \xi_{\bm q})),\\
    K^{(d)}(\bm k) &= \frac{1}{N^2}\sum_{\{\bm q_i\}} \delta(\omega + \varepsilon_{\bm q_1} - \varepsilon_{\bm q_2} - \varepsilon_{\bm k + \bm q_1 - \bm q_2}).
\end{align}
\begin{figure}[t!]
\centering
\includegraphics[width=\linewidth]{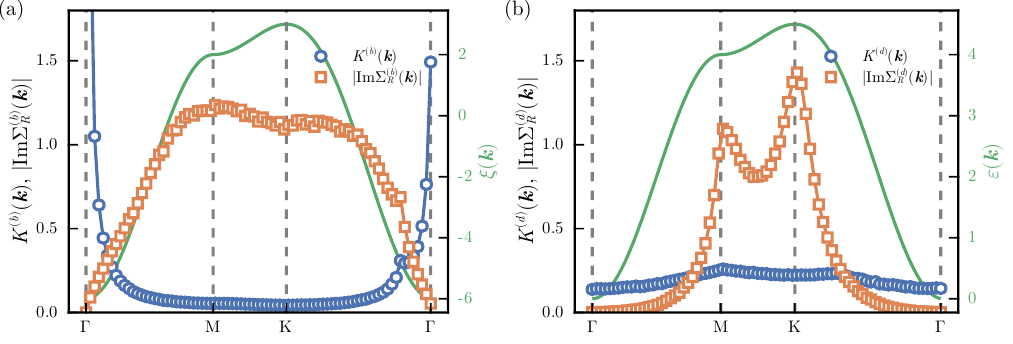}
\caption{Scattering density of states $K$ and imaginary part of the self-energy $\im\Sigma$ due to (a) electron-magnon interactions and (b) magnon-magnon interactions, together with free electron dispersion $\xi(\bm k)$ and free magnon dispersion $\varepsilon(\bm k)$, respectively, along the high symmetry momentum path. Model parameters used: $t/J=1$, $J_K/J=10$ and $T=1$. Error bars of numerical integration are smaller than the symbol size.}
\label{fig3}
\end{figure}

\begin{figure}
\centering
\includegraphics[width=\linewidth]{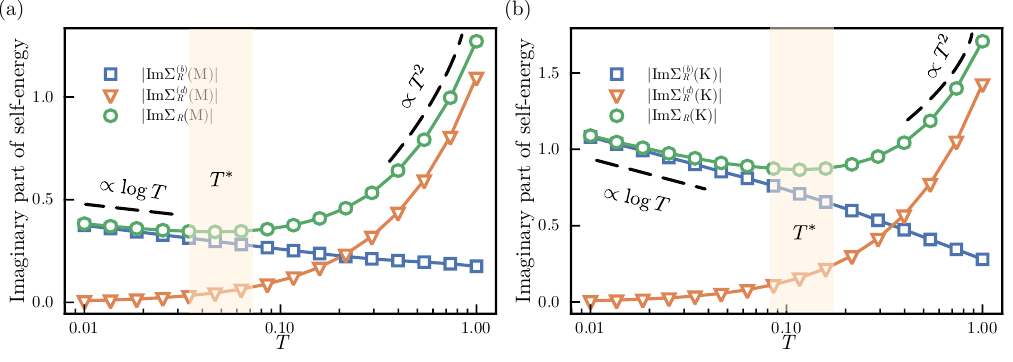}
\caption{Imaginary part of the self-energy from two types of interactions and their total contributions (namely, the magnon damping rate), as a function of temperature $T$, at the M (a) and K (b) point in the first BZ. The shaded rectangle with $T^\ast$ indicates the regime where magnon damping minimum occurs. Model parameters used: $t/J=2$, $J_K/J=12$. Error bars of numerical integration are smaller than the symbol size.}
\label{fig4}
\end{figure}

In Fig.~\ref{fig3}, we show the imaginary part of self-energy $\im \Sigma$ and the corresponding scattering density of states $K$, for each contribution. The multi-dimensional integrations over internal momenta are performed using \texttt{MCIntegration.jl}~\cite{chen2023}. It is found that for the electron-magnon interactions, the scattering density of states is divergent around $\Gamma$ point and suppressed otherwise; while for the magnon-magnon interactions, it peaks at $M$ point and reaches minimum at $\Gamma$ and $K$. For the respective contributions to the magnon damping, namely the imaginary part of the self-energy, both become larger when going away from the BZ center, and become largest near the BZ boundary. This behavior is consistent with the experimental observation in {\fgt} \cite{Bao2023}.

In Fig.~\ref{fig4}, we show the temperature dependence of the damping rate resulting from two types of interactions and their total contribution. In the large and small $T$ limit, imaginary part of Eq.~\eqref{sigmab} and \eqref{sigmad} can be understood analytically: On the one hand, it can be shown straightforwardly that $\im \Sigma_R^{(b)}\propto 1/T$ for $T\rightarrow\infty$ and $\im \Sigma_R^{(d)}$ decay exponentially for $T\rightarrow0$, respectively. On the other hand, $\im \Sigma_R^{(d)}\propto T^2$ in the high-$T$ limit based on a simple series expansion, while $\im \Sigma_R^{(b)}\propto \log T$ in the low-$T$ limit due to Kondo renormalization. Therefore, there is indeed a magnon damping minimum at an intermediate temperature $T^\ast$, inside the shaded regime in Fig.~\ref{fig4}. Using the same model parameters as the ones given in Fig.~3 of the main text with $J_K=10$, we obtain the gray dotted line given there.

\section{Tensor-network and Mean-field Calculations of the Phase Diagram}

\subsection{Quantum critical point between ferromagnetic and Kondo insulator phases}
In this section, we show more detailed calculations on the
quantum critical point (QCP) between ferromagnetic (FM) and Kondo insulator (KI) phases. 
We compute both the ground state and 
low-temperature properties, e.g., the heat capacity, which witness the 
existence of  QCP.

\begin{figure*}[htbp]
  \includegraphics[angle=0,width=0.7\linewidth]{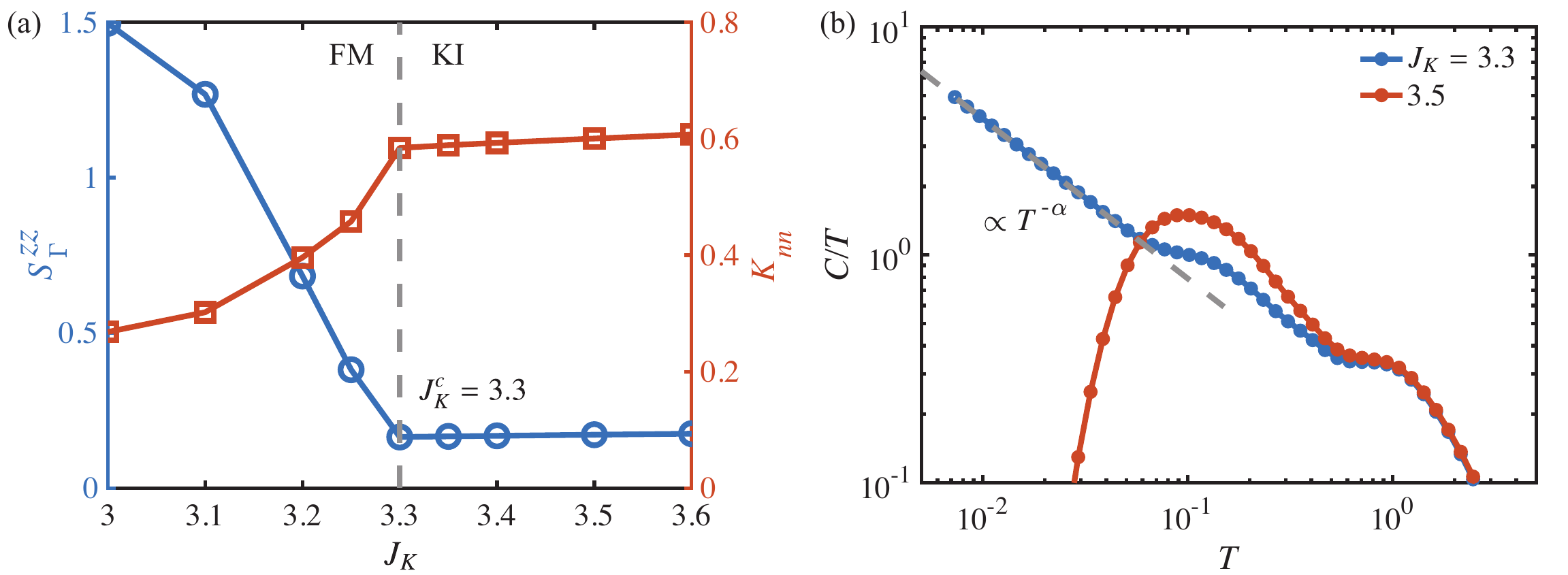}
  \caption{(a) shows the ground-state spin structure at $k=0$ and Kondo correlation function $K_{nn}$, 
as function of Kondo coupling $J_K$, with $J_H=-1$, $t=2$, $N=36\times2$.
(b) shows $C/T$ data with $N=36\times2$ and $J_K = 3.3,~3.5$.
  At the quantum critical point, the lower part of $C/T$ diverges algebraically with $\alpha \sim 0.7$.}
  \label{SFig:FMQCP}
\end{figure*}

To determine the ferromagnetic quantum critical point, we perform DMRG 
simulations on a $N=36\times2$ geometry, with $D = 1600$ bond states
retained and truncation error around $\epsilon \approx 5e^{-7}$. 
We also calculate the 
spin structure factor $S^{zz}_{\Gamma} = \frac{1}{L} \sum_{m,n}  \langle 
S_m^z S_n^z \rangle e^{{\rm i}(n-m)k}$ ($k=0$, $\Gamma$ point) and  local 
Kondo correlation function $K_{nn}=-\frac{1}{L}\sum_n \langle {\bf\it S}_n \cdot 
{\bf\it s}_n\rangle$. As shown in Fig.~\ref{SFig:FMQCP}(a), FM spin correlation decreases fast
when increasing $J_K^c$ and converges to a small value in the KI phase, while the Kondo correlation monotonically
increases and saturates instead.

We further perform XTRG simulation at the 
putative phase transition point $J_K^c = 3.3$. As shown in Fig.~\ref{SFig:FMQCP} (b),
the low temperature heat capacity displays algebraic behavior as $C/T \sim 
T^{-0.7}$, signalling quantum criticality.

\subsection{Determination of the crossover temperature}
\begin{figure*}[htbp]
\includegraphics[angle=0,width=0.7\linewidth]{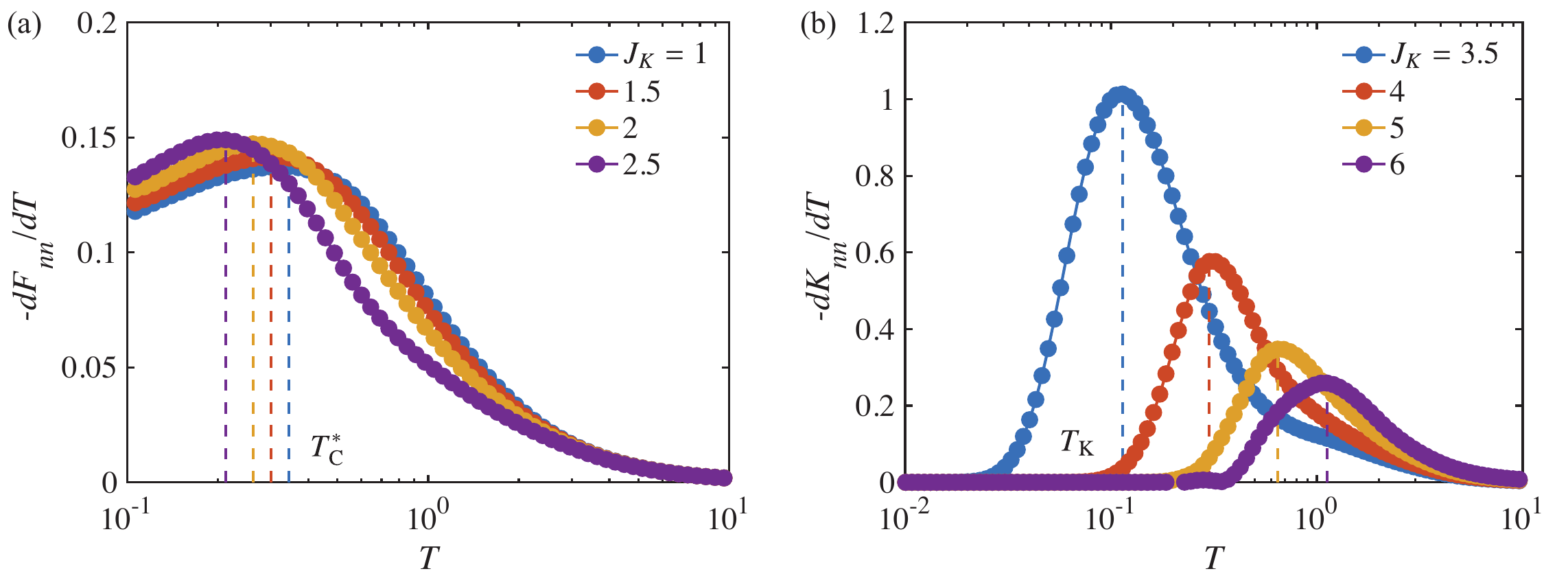}
\caption{(a) shows the -$d F_{nn}/d T$ in the FM phase 
and (b) shows the -$d K_{nn}/d T$ in the KI phase
with $ N=36\times 2$ and $D = 1600$. The dashed line indicates $T_{\rm C}^*$ 
and $T_{\rm K}$ respectively.}
  \label{SFig:C_KI}
\end{figure*}

Although there is no finite-temperature phase transition in 1D quantum system, 
the crossover temperature can be determined form the thermodynamic quantities 
and correlation function. In the FM phase, we obtain the crossover temperature 
$T^*_{\rm C}$ from the  the temperature derivatives of local moment
correlations -$d F_{nn}/dT$ with $F_{nn} = \frac{1}{L-1} \sum_n \langle {\bf S}_n \cdot 
{\bf S}_{n+1} \rangle$. 
The corresponding result is shown in Fig.~\ref{SFig:C_KI}(a). As the Kondo coupling increasing, the 
crossover temperature decreases. In the Kondo insulator phase, we determine the 
crossover temperature by -$d K_{nn}/dT$. As shown in Fig.~\ref{SFig:C_KI}(b), $T_{\rm K}$ 
has positive correlation to the Kondo coupling $J_K$. All the above calculation are 
performed by XTRG with $N=36\times2$ and $D = 1600$ and the corresponding 
result is shown in Fig.~1(a) of the main text.

\subsection{Finite-temperature properties in the FM phase}
\begin{figure*}[htbp]
  \includegraphics[angle=0,width=0.4\linewidth]{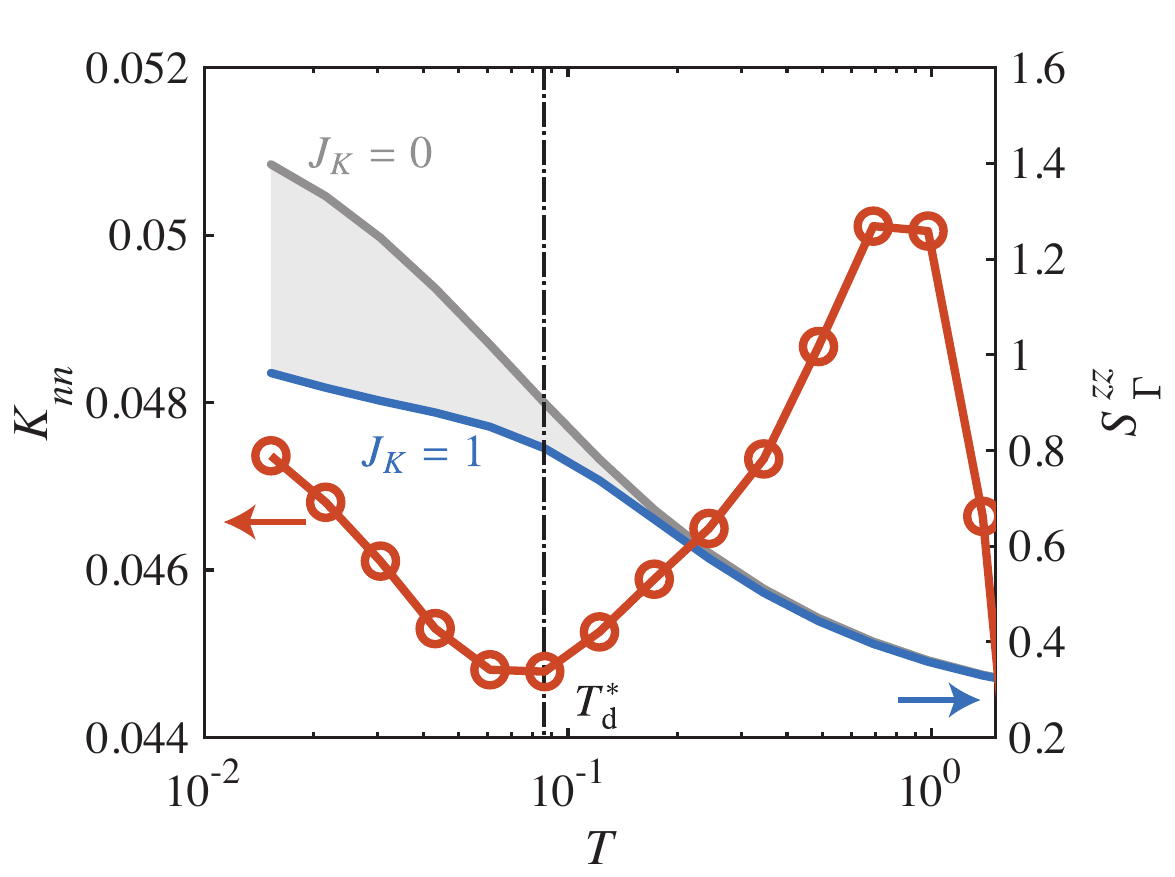}
  \caption{Kondo correlation $K_{nn}$ (red circle) and spin structure factor $S^{zz}_{\Gamma}$
  (blue solid line) with $N=16\times2$, $t=2$, $J_H=-1$, $J_K=1$ and $D=1600$.
  The gray solid line shows $S^{zz}_{\Gamma}$ with $J_K=0$ and the gray shadow region 
  shows the difference between $J_K=0$ and $J_K=1$. The vertical point-solid line 
  shows the minimum damping temperature $T_{\rm d}^*$ determine by $\gamma_{\pi}$ (see
  Fig.~3 in the main text).}
  \label{SFig:FMCorr}
\end{figure*}

Here we discuss the Kondo correlation $K_{nn}$ and FM spin structure factor 
$S^{zz}_{\Gamma}$ in the FM region (with the same model parameter in Fig.~3 of the main text). 
As shown in Fig.~\ref{SFig:FMCorr}, $K_{nn}$ also displays a minimum value around 
$T_{\rm d}^*$, while $S^{zz}_{\Gamma}$ behaves similarly for both $J_K=0$ and $J_K\neq 0$ above $T_{\rm d} ^ *$.
 These results indicate that 
the Kondo coupling $J_K$ prominently changes the thermodynamical properties 
below $T_{\rm d}^*$. Notably, $S^{zz}_{\Gamma}$ keeps increasing as system 
cooling down even below $T_{\rm d} ^ *$ which means that the local moments are not
fully screened by the itinerant electrons in this case.

\subsection{Determination of the mean-field phase diagram}
\begin{figure*}[htbp]
  \includegraphics[angle=0,width=0.35\linewidth]{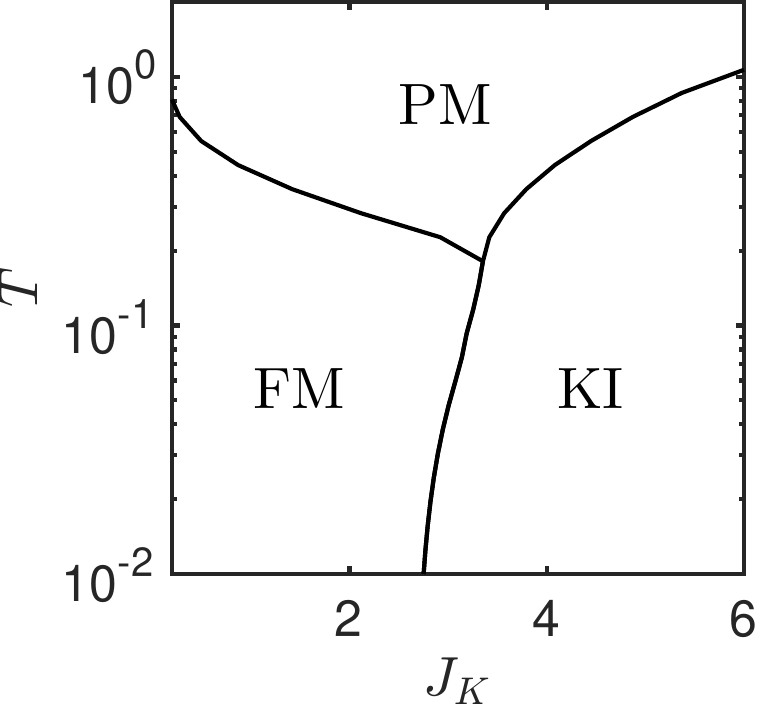}
  \caption{Phase diagram of the 1D FMKH model ($t=2$, $J_H=-1$), based on the effective mean field
  theory by solving Eq.~\eqref{emft_eqns} self-consistently. The FM phase is
  identified with nonzero $m_c$ and $m_f$. The KI phase is identified with
  nonzero $V$.}
  \label{smfig:emft}
\end{figure*}

In this section, we discuss an effective mean-field theory which can treat the FM order and local Kondo screening effect on an equal footing. Hence a complete phase diagram containing both the FM phase at small $J_K$ and KI phase at large $J_K$ can be obtained. The starting point of this effective mean-field theory is to notice that the Kondo coupling term can be decomposed into longitudinal and transversal parts \cite{zhang2000},
\begin{equation}
  \bdsb S_n \cdot \bdsb s_n = S_n^z s_n^z - \frac{1}{4}[(c^\dagger_{n\uparrow} f_{n\uparrow} + f^\dagger_{n\downarrow}c_{n\downarrow})^2 + (c^\dagger_{n\downarrow} f_{n\downarrow} + f^\dagger_{n\uparrow} c_{n\uparrow})^2 ],
  \label{eq:kondocoupling}
\end{equation}
where we have used the pseudo-fermion representation for the local spin $S_n = \frac{1}{2}\sum_{\alpha\beta} f^\dagger_{n\alpha} \bdsb \sigma_{\alpha\beta} f_{n\beta}$, with a local constraint
\begin{equation}
  \sum_\sigma f^\dagger_{n\sigma} f_{n\sigma} = 1,
  \label{eq:lc}
\end{equation}
enforced at each site $n$. Equation \eqref{eq:kondocoupling} inspired one to
define, besides the standard FM order parameters
\begin{equation}
  m_c = \expval{S^z_n},\quad m_f = \expval{s^z_n},
\end{equation}
also the Kondo hybridization order parameter as follows,
\begin{equation}
  V = \expval{c^\dagger_{n\uparrow} f_{n\uparrow} + f^\dagger_{n\downarrow}c_{n\downarrow}}.
\end{equation}
Using these order parameters, one can decouple the original Hamiltonian, $H = H_t + H_H + H_K$ with
\begin{eqnarray}
  H_t &=& -t\sum_n c^\dagger_n c_{n+1} + \hc,\\
  H_H &=& J_H\sum_n \bdsb S_n \cdot \bdsb S_{n+1},\\
  H_K &=& J_K\sum_n \bdsb S_n \cdot \bdsb s_n,
\end{eqnarray}
to a quadratic form. Firstly, the hopping term, $H_t$, is unchanged. Secondly,
the ferromagnetic Heisenberg term becomes
\begin{equation}
  H_H \rightarrow J_H\sum_n m_f (f^\dagger_{n\uparrow} f_{n\uparrow} - f^\dagger_{n\downarrow} f_{n\downarrow}) - J_HL m_f^2.
\end{equation}
And lastly, the Kondo coupling term becomes
\begin{eqnarray}
  H_K & \rightarrow & J_K \sum_{n = 1}^L \left[ \frac{m_c}{2} (f_{n
  \uparrow}^{\dag} f_{n \uparrow} - f^{\dag}_{n \downarrow} f_{n \downarrow})
  + \frac{m_f}{2} (c^{\dag}_{n \uparrow} c_{n \uparrow} - c^{\dag}_{n
  \downarrow} c_{n \downarrow}) - m_c m_f \right]\nonumber\\
  &  & + J_K \sum_{n = 1}^L \left[ - \frac{1}{2} V (c^{\dag}_{n \uparrow} f_{n
  \uparrow} + f^{\dag}_{n \downarrow} c_{n \downarrow}) - \frac{1}{2} V
  (c^{\dag}_{n \downarrow} f_{n \downarrow} + f^{\dag}_{n \uparrow} c_{i
  \uparrow}) + \frac{1}{2} V^2 \right] .
\end{eqnarray}
We also introduce a Lagrangian multiplier $\lambda$ to enforce the local
constraint Eq.~\eqref{eq:lc},
\begin{equation}
  \lambda \sum_{n = 1}^L (f^{\dag}_{n \sigma} f_{n \sigma} - 1),
\end{equation}
and a chemical potential $\mu$ to fix the total $c$ fermion number at half
filling,
\begin{equation}
  - \mu \sum_{n = 1}^L (c^{\dag}_n c_n - 1) . 
\end{equation}
Hence the mean-field Hamiltonian in momentum space can be written in a compact
form
\begin{equation}
  H_{\text{MF}} = \sum_{k, \sigma} (c^{\dag}_{k \sigma}, f^{\dag}_{k \sigma})
   \left(\begin{array}{cc}
     \varepsilon_{k \sigma} & - \frac{1}{2} J_KV\\
     - \frac{1}{2} J_KV & \lambda_{\sigma}
   \end{array}\right) \left(\begin{array}{c}
     c_{k \sigma}\\
     f_{k \sigma}
   \end{array}\right) + L \varepsilon_0,
\end{equation}
where
\begin{eqnarray}
  \varepsilon_{k \sigma} & = & - 2 t \cos k + \frac{J_Km_f}{2} \sigma - \mu,\\
  \lambda_{\sigma} & = & \lambda + \left( \frac{J_Km_c}{2} + J_Hm_f \right)
  \sigma,\\
  \varepsilon_0 & = & \frac{J_KV^2}{2} - J_Km_c m_f - \lambda + \mu - J_Hm_f^2 .
\end{eqnarray}
The quasiparticle excitation spectra are thus obtained by
\begin{eqnarray}
  E_{k \sigma l} & = & \frac{1}{2} \left[ \varepsilon_{k \sigma} +
  \lambda_{\sigma} + l \sqrt{(\varepsilon_{k \sigma} - \lambda_{\sigma})^2 +
  (J_KV)^2} \right],
\end{eqnarray}
which corresponds to four quasiparticle bands, with $\sigma = \pm$ and $l =
\pm$. The corresponding Helmholtz free energy density reads
\begin{equation}
  f = - \frac{1}{\beta L} \sum_k \sum_{l, \sigma = \pm} \log [1 + \exp (-
   \beta E_{k \sigma l})] + \varepsilon_0 .
\end{equation}
which should be minimized with respect to the five unknown parameters:
\begin{subequations}
\begin{align}
  0 & =  \partial_{m_f} f\nonumber\\
  & =  \frac{1}{L} \sum_k \sum_{l, \sigma = \pm} \frac{e^{- \beta E_{k
  \sigma l}}}{1 + \exp (- \beta E_{k \sigma l})} \partial_{m_f} E_{k \sigma l}
  - J_Km_c - 2 J_Hm_f\nonumber\\
  & =  \frac{1}{L} \sum_k \sum_{l, \sigma = \pm} n(E_{k\sigma l}) \sigma \left[  \frac{J_H}{2} + 
  \frac{J_K}{4} + l \left(
  \frac{J_K}{4}-\frac{J_H}{2} \right) \frac{\varepsilon_{k \sigma} -
  \lambda_{\sigma}}{\sqrt{(\varepsilon_{k \sigma} - \lambda_{\sigma})^2 +
  (KV)^2}} \right] - J_Km_c - 2 J_Hm_f\\
  0 & =  \partial_{m_c} f\nonumber\\
  & =  \frac{1}{L} \sum_k \sum_{l, \sigma = \pm} n(E_{k\sigma l}) \frac{J_K \sigma}{4} \left[ 1 - l \frac{\varepsilon_{k \sigma}
  - \lambda_{\sigma}}{\sqrt{(\varepsilon_{k \sigma} - \lambda_{\sigma})^2 +
  (J_KV)^2}} \right] - J_Km_f\\
  0 & =  \partial_V f\nonumber\\
  & =  \frac{1}{L} \sum_k \sum_{l, \sigma = \pm} n(E_{k\sigma l}) \frac{lJ_K^2 V / 2}{\sqrt{(\varepsilon_{k \sigma} -
  \lambda_{\sigma})^2 + (J_KV)^2}} + J_KV\\
  0 & =  \partial_{\lambda} f\nonumber\\
  & =  \frac{1}{L} \sum_k \sum_{l, \sigma = \pm} n(E_{k\sigma l}) \frac{1}{2} \left[ 1 - l \frac{\varepsilon_{k \sigma} -
  \lambda_{\sigma}}{\sqrt{(\varepsilon_{k \sigma} - \lambda_{\sigma})^2 +
  (J_KV)^2}} \right] - 1\\
  0 & =  \partial_{\mu} f\nonumber\\
  & =  \frac{1}{L} \sum_k \sum_{l, \sigma = \pm} n(E_{k\sigma l}) \frac{1}{2} \left[ - 1 - l \frac{\varepsilon_{k \sigma} -
  \lambda_{\sigma}}{\sqrt{(\varepsilon_{k \sigma} - \lambda_{\sigma})^2 +
  (J_KV)^2}} \right] + 1,
\end{align}
\label{emft_eqns}
\end{subequations}
with $n(E_{k\sigma l}) = (e^{\beta E_{k \sigma l}} + 1)^{-1}$ being the Fermi
function. We numerically solve above five equations self-consistently. The
resulting finite temperature phase diagram for $(J_H,t) =(-1,2)$ is shown in
Fig.~\ref{smfig:emft}.

\section{Excitation Spectrum in the Strong-coupling Limit: Bond-operator Theory}

In this appendix, we give a strong-coupling mean-field analysis of the FMKH
model for $K\gg J,t$, using bond-operator theory. This approach was originally proposed by
Sachdev and Bhatt for the dimerized spin systems \cite{sachdev1990}. Its
generalization to the Kondo lattice systems was firstly given by Jurecka and
Brenig \cite{jurecka2001}. R.~Eder recently gave an improved treatment to
obtained both the fermionic quasiparticle and spin excitation spectrum for the
Kondo lattice \cite{eder2019}. We here generalize it to the FMKH model.

\subsection{Bond-operator basis}
The Kondo-coupling term leads to eight eigenstates, defined as follows \cite{jurecka2001}:
\begin{subequations}
  \begin{align}
    \ket{s} &= s^\dagger\ket{0} = \frac{1}{\sqrt 2}(f^\dagger_\downarrow  c^\dagger_\uparrow + c^\dagger_\downarrow f^\dagger_\uparrow) \ket{0},\\
    \ket{t_x} &= t^\dagger_x\ket{0} = \frac{-1}{\sqrt{2}} (c^\dagger_\uparrow f^\dagger_\uparrow - c^\dagger_\downarrow f^\dagger_\downarrow) \ket{0},\\
    \ket{t_y} &= t^\dagger_y\ket{0} = \frac{\rm i}{\sqrt{2}} (c^\dagger_\uparrow f^\dagger_\uparrow + c^\dagger_\downarrow f^\dagger_\downarrow) \ket{0},\\
    \ket{t_z} &= t^\dagger_z\ket{0} = \frac{1}{\sqrt{2}} (c^\dagger_\uparrow f^\dagger_\downarrow + c^\dagger_\downarrow f^\dagger_\uparrow) \ket{0},\\
    \ket{a_\sigma} &= a^\dagger_\sigma \ket{0} = f^\dagger_\sigma \ket{0},\\
    \ket{b_\sigma} &= b^\dagger_\sigma \ket{0} = c^\dagger_\uparrow c^\dagger_\downarrow f^\dagger_\sigma \ket{0}.
  \end{align}
  \label{bobasis}
\end{subequations}
Note the $f$ fermion has to satisfy the local constraint at each site,
\begin{equation}
  f^\dagger_{n\sigma} f_{n\sigma} = 1,\quad \forall \ n.
  \label{eq:nf}
\end{equation}
Hereafter Einstein summation rule over repeated Greek alphabet (i.e., spin
indices) is assumed. When the spin index is omitted, the corresponding operators
should be viewed as a two-component object. The first four particles in
Eq.~\eqref{bobasis} are bosons; and the rest are fermions. They together satisfy
the local constraint
\begin{equation}
  s^\dagger_n s_n + a^\dagger_{n\sigma} a_{n\sigma} + b^\dagger_{n\sigma} b_{n\sigma} + \bdsb{t}^\dagger_n \cdot \bdsb{t}_{n} = 1,\quad \forall n
  \label{localc1}
\end{equation}
which ensures Eq.~\eqref{eq:nf} automatically.

\subsection{Operators in bond-operator representation}
Here we give explicit form of relevant operators in this representation. First of all, the spin operator for itinerant electrons and local moments
becomes, respectively,
\begin{align}
  s^\sigma_n &= \frac{1}{2} (-t^\dagger_{n\sigma} s_n - s^\dagger_{n} t_{n\sigma} -{\rm i}\epsilon_{\sigma\alpha\beta}t^\dagger_{n\alpha}t_{n\beta}),\\
  S^\sigma_n &= \frac{1}{2} (t^\dagger_{n\sigma} s_n + s^\dagger_n t_{n\sigma} - {\rm i}\epsilon_{\sigma\alpha\beta} t^\dagger_{n\alpha} t_{n\beta} ) +S^\sigma_{a,n} + S^\sigma_{b,n},
\end{align}
where $S_{a,n}^\mu = \frac{1}{2}a^\dagger_n \sigma^\mu a_n$ and $S_{b,n}^\mu = \frac{1}{2}b^\dagger_n\sigma^\mu b_n$. Then the Kondo coupling term can be shown to be
\begin{equation}
  J_K \sum_n {\bf S}_n \cdot {\bdsb s}_n = \sum_n [\frac{3J_K}{4} (b^\dagger_{n\sigma} b_{n\sigma} + a^\dagger_{n\sigma} a_{n\sigma}) + J_K \bdsb{t}^\dagger_{n}\cdot\bdsb{t}_n] - \frac{3J_NK}{4},
\end{equation}
which is diagonal as expected. Next, the itinerant electron annihilation operator becomes
\begin{equation}
  c_n = \frac{1}{\sqrt 2}[(-s_n + \bdsb t_n \cdot \bdsb \tau) {\rm i}\tau^y a^\dagger_n + (s^\dagger_n +\bdsb t^\dagger_n \cdot \bdsb \tau) b_n],
  \label{cn}
\end{equation}
from which we can obtain the hopping term for itinerant electrons,
\begin{equation}
  -t \sum_{n} c^\dagger_n c_{n+1} + \hc = H_{c1} + H_{c2} + H_{c3} + H_{c4},
\end{equation}
with
\begin{align}
  H_{c1} &= \frac{1}{2} \sum_{n,m} t_{n,m}[(-s^\dagger_n s_m a^\dagger_m a_n + s_n s^\dagger_m b^\dagger_n b_m - (s_n s_m b^\dagger_n {\rm i}\tau^y a^\dagger_m+ \hc)],\label{h1a}\\
  H_{c2} &= \frac{1}{2} \sum_{n,m} t_{n,m}[ -\bdsb t^\dagger_n \cdot \bdsb t_m a^\dagger_m a_n + \bdsb t_n \cdot \bdsb t^\dagger_m b^\dagger_n b_m + (\bdsb t_n \cdot \bdsb t_m b^\dagger_n {\rm i}\tau^y a^\dagger_m + \hc)], \label{h2a}\\
  H_{c3} &= \frac{1}{2}\sum_{n,m} t_{n,m} \{(s^\dagger_n \bdsb t_m + s_m \bdsb t^\dagger_n) \cdot (\bdsb a_{m,n} + \bdsb b_{m,n}) + [(s_m \bdsb t_n - s_n \bdsb t_m) \cdot \bdsb \pi^\dagger_{n,m} + \hc]\},\\
  H_{c4} &= \frac{1}{2}\sum_{n,m} t_{n,m} [-\bdsb t^\dagger_n \times \bdsb t_m \cdot {\rm i}(\bdsb a_{m,n} + \bdsb b_{m,n}) + (\bdsb t_m \times \bdsb t_n \cdot {\rm i}\bdsb \pi^\dagger_{m,n} + \hc)],
\end{align}
Here $t_{m,n} = -t$ for the nearest-neighbors, and vanishing otherwise.
And the following vectorized operators have been used: $\bdsb a_{m,n} = a^\dagger_m \bdsb \tau b_n$, $\bdsb b_{m,n} = b^\dagger_{m} \bdsb \tau b_{n}$ and $\bdsb \pi_{m,n}^\dagger = b^\dagger_m \bdsb \tau {\rm i}\tau^y a^\dagger_n$. Interestingly, the hopping term, which is originally quadratic, now becomes quartic.
Such behavior is common for strong-coupling approaches, see also \cite{wang2021}.
Lastly, the Heisenberg coupling term for localized spins becomes
\begin{equation}
  J_H \sum_n {\bf S}_n \cdot {\bf S}_{n+1} = H_{s1} + H_{s2} + H_{s3} + H_{s4},\quad J>0,
\end{equation}
where
\begin{equation}
\begin{split}
  H_{s1} &= \frac{1}{8} \sum_{n,m}J_{n,m}  (\bdsb t^\dagger_n s_n + s^\dagger_n \bdsb t_n) \cdot (\bdsb t^\dagger_m s_m + s^\dagger_m \bdsb t_m),\\
  H_{s2} &= -\frac{\rm i}{4} \sum_{n,m} J_{n,m} (\bdsb t^\dagger_n s_n + s^\dagger_n \bdsb t_n) \cdot \bdsb t^\dagger_m \times \bdsb t_m,\\
  H_{s3} &= -\frac{1}{8} \sum_{n,m} J_{n,m} [(\bdsb t^\dagger_n \cdot \bdsb t^\dagger_m) (\bdsb t_n \cdot \bdsb t_m) - t^\dagger_{n\alpha}t_{n\beta} t^\dagger_{m\beta} t_{m\alpha}],\\
  H_{s4} &=\frac{1}{2} \sum_{n,m} J_{n,m} [(\bdsb t^\dagger_n s_n + s^\dagger_n \bdsb t_n - i\bdsb t^\dagger_n \times \bdsb t_n) \cdot (\bdsb S_{a,m} + \bdsb S_{b,m}) + (\bdsb S_{a,n} + \bdsb S_{b,n}) \cdot (\bdsb S_{a,m} + \bdsb S_{b,m})],
  \end{split}
\end{equation}
with $J_{n,m}=J_H<0$ for nearest neighbors and vanishing otherwise.

\subsection{Simplest mean-field approximation}

In the simplest mean-field approach, we condense singlets,
$s^\dagger_n, s_n\rightarrow s$, and collect terms up to quadratic order in the remaining operators.
The Kondo coupling term is unchanged,
\begin{align}
  J_K\sum_n \bdsb S_n \cdot \bdsb s_n &\rightarrow \sum_n [\frac{3J_K}{4} (b^\dagger_{n\sigma} b_{n\sigma} + a^\dagger_{n\sigma} a_{n\sigma}) + J_K \bdsb{t}^\dagger_{n}\cdot\bdsb{t}_n] - \frac{3NJ_K}{4}\nonumber\\
  &= \sum_k [\frac{3J_K}{4} (b^\dagger_k b_k + a^\dagger_k a_k) + J_K \bdsb t^\dagger_k\cdot \bdsb t_k] - \frac{3NJ_K}{4}.
\end{align}
The hopping term for itinerant electrons becomes
\begin{align}
  -t \sum_{n} c^\dagger_n c_{n+1} + \hc  &\rightarrow \frac{s^2}{2} \sum_{n,m} t_{n,m}[(-a^\dagger_m a_n +b^\dagger_n b_m - (b^\dagger_n {\rm i}\tau^y a^\dagger_m+ \hc)]\nonumber\\
  &=- ts^2 \sum_k [\cos k (-a^\dagger_k a_k + b^\dagger_k b_k) - (\cos k b^\dagger_k {\rm i}\tau^y a^\dagger_{-k} + \hc) ].
\end{align}
And finally the Heisenberg coupling term becomes
\begin{align}
  J_H\sum_{n} \bdsb S_n \cdot \bdsb S_{n+1} &\rightarrow \frac{s^2}{8} \sum_{n,m}J_{n,m}  (\bdsb t^\dagger_n + \bdsb t_n) \cdot (\bdsb t^\dagger_m + \bdsb t_m)\nonumber\\
  &= \frac{s^2J_H}{4}\sum_k [2\cos k \bdsb t^\dagger_k \cdot \bdsb t_k + (\cos k \bdsb t^\dagger_k \cdot \bdsb t^\dagger_{-k} + \hc)].
\end{align}
We further introduce two Lagrange multipliers, $\lambda$ and $\mu$, to enforce Eq.~\eqref{localc1} globally,
\begin{equation}
  -\lambda \sum_n \left[ s^\dagger_n s_n + \bdsb t^\dagger_n \cdot \bdsb t_n + a^\dagger_n a_n + b^\dagger_n b_n - 1 \right]
\end{equation}
and also to fix the $c$ fermion number at half-filling,
\begin{align}
  0 &= -\mu \sum_n\left( s^\dagger_n s_n + \bdsb t^\dagger_n\cdot \bdsb t_n + 2b^\dagger_{n} b_{n} - 1\right)\nonumber\\
  &= -\mu \sum_n \left( b^\dagger_n b_n - a^\dagger_n a_n \right).
\end{align}
Thus the resulting mean-field Hamiltonian reads
\begin{equation}
  H_{\mf} = H_F + H_B -\frac{3NJ_K}{4} + N(1-s^2) \lambda ,\label{hmf}
\end{equation}
where
\begin{align}
  H_F &= \sum_k [ (\frac{3J_K}{4} - \lambda + \mu + s^2 t \cos k) a^\dagger_k a_k
  + (\frac{3J_K}{4} - \lambda - \mu - s^2 t\cos k) b^\dagger_k b_k 
  + (s^2 t\cos k b^\dagger_k {\rm i}\tau^y a^\dagger_{-k} + \hc)],\label{eq:hf}\\
  H_B &= \sum_k [ ( J_K - \lambda - \frac{s^2 J }{2} \cos k) \bdsb t^\dagger_k \cdot \bdsb t_k 
  + (\frac{s^2J_H}{4} \cos k \bdsb t^\dagger_k \cdot \bdsb t^\dagger_{-k} + \hc)].\label{eq:hb}
\end{align}

\subsection{Solving the fermionic part}
The fermionic sector, Eq.~\eqref{eq:hf}, can be written as
\begin{equation}
  H_F = \sum_k \begin{pmatrix}
    a^\dagger_k & b^\dagger_k
  \end{pmatrix} A_k \begin{pmatrix}
    a_k\\
    b_k
  \end{pmatrix} + \left[\frac{1}{2}\begin{pmatrix}
    a^\dagger_k & b^\dagger_k
  \end{pmatrix} B_k \begin{pmatrix}
    a^\dagger_{-k}\\
    b^\dagger_{-k}
  \end{pmatrix}+\hc \right],
  \label{eq:hf1}
\end{equation}
where
\begin{align}
  A_k &= \begin{pmatrix}
    \frac{3J_K}{4} - \lambda + \mu + s^2t\cos k & 0\\
    0 & \frac{3J_K}{4} - \lambda - \mu - s^2 t \cos k
  \end{pmatrix},\label{eq:AFK}\\
  B_k &= \begin{pmatrix}
    0 & s^2t\cos k {\rm i}\tau^y\\
    s^2t\cos k{\rm i}\tau^y & 0
  \end{pmatrix}.
\end{align}
By defining the Nambu spinor $\psi^\dagger_k = \begin{pmatrix}
  a^\dagger_k & b^\dagger_k & a_{-k} & b_{-k}
\end{pmatrix}$,
Eq.~\eqref{eq:hf1} can be further recast into the Bogoliubov-de Gennes (BdG) form, $H_F = \frac{1}{2} \sum_k \psi^\dagger_k M_k \psi_k + \frac{1}{2}\sum_k \Tr {A_k}$, where the BdG matrix reads $M_k = \begin{pmatrix}
    A_k & B_k \\
    B^\dagger_k & -A^T_{-k}
  \end{pmatrix}$. This matrix can be diagonalized by using a unitary matrix $U_k$,
such that $\psi_k = U_k \varphi_k$, and
\begin{align}
  \frac{1}{2}\sum_k \psi^\dagger_k M_k \psi_k &= \frac{1}{2}\sum_k \varphi^\dagger_k D_k \varphi_k\nonumber\\
  &= \sum_k e_{a,k} \tilde a^\dagger_k \tilde a_k + e_{b,k} \tilde b_k \tilde b_k - \sum_k \sum_{l=a,b} e_{l,k}.
\end{align}
Note that there is a factor of two for the last term, which is due to double
degeneracy. Here $\varphi^\dagger_k = \begin{pmatrix} \tilde a^\dagger_k &
\tilde b^\dagger_k & \tilde a_k & \tilde b_k \end{pmatrix}$ is the Nambu spinor
form of the fermionic quasiparticle, $D_k$ is a diagonal matrix, and the
eigenvalues are
\begin{equation}
  e_{a(b), k} = \sqrt{(\frac{3J_K}{4}-\lambda)^2 + s^4 t^2 \cos^2 k } \pm \mu \pm s^2 t \cos k.
  \label{eq:fdisp}
\end{equation}
Hence Eq.~\eqref{eq:hf} becomes
\begin{equation}
  H_F = \sum_k e_{a,k} \tilde a_{k}^\dagger \tilde a_k + e_{b,k} \tilde b^\dagger_k \tilde b_k - \sum_k \sum_{l=a,b} e_{l,k} + \frac{1}{2}\sum_k \Tr{A_k}.
\end{equation}

\subsection{Solving the bosonic part}
The bosonic sector, Eq.~\eqref{eq:hb} can be further written as
\begin{equation}
  H_B = \sum_k \bdsb t^\dagger_k A_k \bdsb t_k + \left(\frac{1}{2}\bdsb t^\dagger_k B_k \bdsb t^\dagger_{-k} + \hc \right),
  \label{eq:hb1}
\end{equation}
where 
\begin{equation}
A_k = K-\lambda + \frac{s^2J_H}{2}\cos k,
\label{eq:ABK}
\end{equation}
and $B_k = \frac{s^2J_H}{2}\cos k$. By defining the Nambu spinor $\psi^\dagger_k = \begin{pmatrix}
  \bdsb t^\dagger_k & \bdsb t_{-k}
\end{pmatrix}$,
Eq.~\eqref{eq:hb1} can be written in the BdG form, $H_b = \frac{1}{2}\sum_k \psi^\dagger_k M_k \psi_k - \sum_k \frac{3}{2}\Tr{A_k}$. Note that there is a factor of $3$ for the last term, which is due to triple
degeneracy. And the BdG matrix reads $M_k = \begin{pmatrix}
    A_k & B_k \\
    B^\dagger_k & A^T_{-k}
  \end{pmatrix}$. This matrix is diagonalized by a pseudounitary matrix $U_k = \begin{pmatrix}
    \cosh \theta_k & \sinh\theta_k\\
    \sinh\theta_k & \cosh\theta_k
  \end{pmatrix}$, with $\tanh 2\theta_k = -\frac{s^2J_H}{2}\cos k/(J_K-\lambda +\frac{s^2J_H}{2}\cos k)$, such that $\psi_k = U_k \varphi_k$, and
\begin{equation}
  \frac{1}{2}\sum_k \psi^\dagger_k M_k \psi_k = \frac{1}{2}\sum_k \varphi^\dagger_k \begin{pmatrix}
    \omega_k & \\
    & \omega_k 
  \end{pmatrix}\varphi_k = \sum_k \omega_k \tilde{\bdsb t}_k \tilde{\bdsb t}_k + \frac{3}{2}\sum_k \omega_k.
\end{equation}
Note again the factor of 3 in the last term.
Here $\varphi^\dagger_k = \begin{pmatrix}
  \tilde{\bdsb t}^\dagger_k & \tilde{\bdsb t}_{-k}
\end{pmatrix}$ is the Nanbu spinor form of the triplon,
with dispersion
\begin{equation}
  \omega_k = \sqrt{(J_K-\lambda)^2 + (J_K-\lambda)s^2 J_H \cos k},
\end{equation}
Hence Eq.~\eqref{eq:hb} becomes
\begin{equation}
  H_b = \sum_k \omega_k \tilde{\bdsb t}^\dagger_k \tilde{\bdsb t}_k + \frac{3}{2}\sum_k \omega_k - \frac{3}{2}\sum_k \Tr{A_k}.
\end{equation}

\subsection{Helmholtz free energy and its minimization}

Based on above analysis,
we can now obtain the Helmholtz free energy density of the system,
\begin{align}
  f &= -\frac{2}{\beta N}\sum_k \sum_{l=a,b} \log[1+\exp(-\beta e_{l,k})]
  + \frac{3}{\beta N} \log[1-\exp(-\beta \omega_{k})]\nonumber\\
  &\quad - \frac{1}{N}\sum_k \sum_{l=a,b}e_{l,k} + \frac{1}{2N}\sum_k\Tr{A_{F,k}})+ \frac{3}{2N}\sum_k \omega_k - \frac{3}{2N}\sum_k\Tr{A_{B,k}}\\
  &\quad -\frac{3J_K}{4} + \lambda (1-s^2),
\end{align}
where $A_{B, k}$ and $A_{F,k}$ are given by Eq.~\eqref{eq:ABK} and
\eqref{eq:AFK}, respectively.

There are three unknown parameters: $s$, $\lambda$ and $\mu$,
which have to be determined by minimizing the free energy density.
For $\lambda$, we have
\begin{align}
  0 &= \partial_\lambda F\\
  &= \frac{3}{N}\sum_k \left[f_B(\omega_k) + \frac{1}{2}\right] 
  \frac{-J_K+\lambda-\frac{s^2J_H}{2}\cos k}{\sqrt{(J_K-\lambda)^2 + (J_K-\lambda) s^2 J_H \cos k}}\nonumber\\
  &\quad - \frac{2}{N}\sum_k \sum_{l=a,b} \left[f_F(e_{l,k}) - \frac{1}{2}\right] \frac{\frac{3J_K}{4} - \lambda}{\sqrt{(\frac{3J_K}{4}-\lambda)^2 + s^4 t^2 \cos^2 k}}\nonumber\\
  &\quad +\frac{1}{2} - s^2.\label{eq:cpdeq1}
\end{align}
For $\mu$, we have
\begin{equation}
	0 = \partial_\mu f = \frac{2}{N}\sum_k [f_F(e_{a,k}) - f_F(e_{b,k})].\label{eq:cpdeq2}
\end{equation}
And for $s$, we have
\begin{align}
	0 &= \partial_s F\\
	&= -\frac{3}{N}\sum_k \left[f_B(\omega_k) + \frac{1}{2}\right] 
	\frac{-(J_K-\lambda) s J_H \cos k}{\sqrt{(J_K-\lambda)^2 + (J_K-\lambda)s^2 J_H \cos k}}\nonumber\\
	&\quad + \frac{2}{N}\sum_k \left[f_F(e_{a,k}) - \frac{1}{2}\right]\left[
	\frac{2 s^3 t^2\cos^2 k}{ \sqrt{(\frac{3J_K}{4}-\lambda)^2 + s^4 t^2 \cos^2 k} }+ 2st\cos k \right]\nonumber\\
	&\quad + \frac{2}{N}\sum_k \left[f_F(e_{a,k}) - \frac{1}{2}\right]\left[
	\frac{2 s^3 t^2 \cos^2 k}{ \sqrt{(\frac{3J_K}{4}-\lambda)^2 + s^4 t^2 \cos^2 k} } - 2st\cos k \right]\nonumber\\
	&\quad -2\lambda s.\label{eq:cpdeq3}
\end{align}
Here $f_{F(B)}(x) = (e^{\beta x}\pm1)^{-1}$ is the Fermi (Bose) function.

\begin{figure*}[!t]
\includegraphics[angle=0,width=0.3\linewidth]{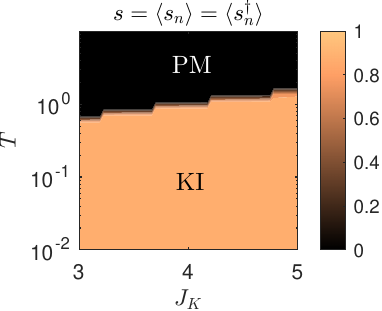}
\caption{Finite-temperature phase diagram for large $J_K$, obtained by
bond-operator theory, with other parameters chosen the same as Fig.~1(a).}
\label{smfig:4}
\end{figure*}

We numerically solve these three coupled nonlinear equations, \eqref{eq:cpdeq1}, 
\eqref{eq:cpdeq2} and \eqref{eq:cpdeq3}. The finite-temperature crossover from 
paramagnetic (PM) phase to the Kondo insulating (KI) phase can be characterized 
by the onset of a nonzero $s$. In Fig.~\ref{smfig:4}, we
plot the condensed singlet $\langle s\rangle$ as a function of temperature $T$ and 
the Kondo coupling $J_K$. It is found
that this strong coupling approach is in excellent agreement with the XTRG result for large $J_K$.

\end{document}